\documentclass[%
superscriptaddress,
amsmath,amssymb,
aip,
reprint,
]{revtex4-1}

\usepackage{hyperref}
\usepackage{graphicx}
\usepackage[dvipsnames]{xcolor}
\usepackage{braket}
\usepackage{dcolumn}% Align table columns on decimal point
\usepackage[version=3]{mhchem}
\usepackage{xcolor}
\usepackage[normalem]{ulem}
\newcounter{pabh}
\newcounter{pabhtodo}
\newcounter{ephraim}
\newcounter{anastasia}
\newcounter{steven}
\newcounter{rob}
\newcounter{alex}

\colorlet{blue}{black}
\colorlet{purple}{black}

\newcommand{\eeff}{$E_{\mathrm{eff}}~$}
\newcommand{\eeffm}{E_{\mathrm{eff}}}

\begin{document}

%% Title and author
\title{Systematic study and uncertainty evaluation of $P,T$-odd \textcolor{blue}{molecular} enhancement factors in BaF} 

\author{Pi A. B. Haase}
\affiliation{Van Swinderen Institute for Particle Physics and Gravity, University of Groningen, 9747 AG Groningen, The Netherlands}
\affiliation{Nikhef, National Institute for Subatomic Physics, 1098 XG Amsterdam, The Netherlands}

\author{Diewertje J. Doeglas}
\affiliation{Van Swinderen Institute for Particle Physics and Gravity, University of Groningen, 9747 AG Groningen, The Netherlands}

\author{Alexander Boeschoten}
\affiliation{Van Swinderen Institute for Particle Physics and Gravity, University of Groningen, 9747 AG Groningen, The Netherlands}
\affiliation{Nikhef, National Institute for Subatomic Physics, 1098 XG Amsterdam, The Netherlands}

\author{Ephraim Eliav}
\affiliation{School of Chemistry, Tel Aviv University, 69978 Tel Aviv, Israel}

\author{Miroslav Ilia{\v{s}}} 
\affiliation{Department of Chemistry, Faculty of Natural Sciences, Matej Bel University, Tajovsk{\'e}ho 40, 97401 Bansk{\'a} Bystrica, Slovakia}

\author{and the eEDM collaboration:}

\author{Parul Aggarwal}
\affiliation{Van Swinderen Institute for Particle Physics and Gravity, University of Groningen, 9747 AG Groningen, The Netherlands}
\affiliation{Nikhef, National Institute for Subatomic Physics, 1098 XG Amsterdam, The Netherlands}

\author{H. L. Bethlem} 
\affiliation{Van Swinderen Institute for Particle Physics and Gravity, University of Groningen, 9747 AG Groningen, The Netherlands}
\affiliation{Department of Physics and Astronomy, VU University Amsterdam, 1081 HV Amsterdam, The Netherlands}

\author{Anastasia Borschevsky}
\email{a.borschevsky@rug.nl}
\affiliation{Van Swinderen Institute for Particle Physics and Gravity, University of Groningen, 9747 AG Groningen, The Netherlands}
\affiliation{Nikhef, National Institute for Subatomic Physics, 1098 XG Amsterdam, The Netherlands}

\author{Kevin Esajas}
\affiliation{Van Swinderen Institute for Particle Physics and Gravity, University of Groningen, 9747 AG Groningen, The Netherlands}
\affiliation{Nikhef, National Institute for Subatomic Physics, 1098 XG Amsterdam, The Netherlands}

\author{Yongliang Hao}
\altaffiliation[Current address: ]{School of Physics and Electronic Engineering, Jiangsu University, Zhenjiang 212013, Jiangsu, China}
\affiliation{Van Swinderen Institute for Particle Physics and Gravity, University of Groningen, 9747 AG Groningen, The Netherlands}
\affiliation{Nikhef, National Institute for Subatomic Physics, 1098 XG Amsterdam, The Netherlands}

\author{Steven Hoekstra}
\affiliation{Van Swinderen Institute for Particle Physics and Gravity, University of Groningen, 9747 AG Groningen, The Netherlands}
\affiliation{Nikhef, National Institute for Subatomic Physics, 1098 XG Amsterdam, The Netherlands}

\author{Virginia R. Marshall}
\affiliation{Van Swinderen Institute for Particle Physics and Gravity, University of Groningen, 9747 AG Groningen, The Netherlands}
\affiliation{Nikhef, National Institute for Subatomic Physics, 1098 XG Amsterdam, The Netherlands}

\author{Thomas B. Meijknecht}
\affiliation{Van Swinderen Institute for Particle Physics and Gravity, University of Groningen, 9747 AG Groningen, The Netherlands}
\affiliation{Nikhef, National Institute for Subatomic Physics, 1098 XG Amsterdam, The Netherlands}

\author{Maarten C. Mooij}
\affiliation{Nikhef, National Institute for Subatomic Physics, 1098 XG Amsterdam, The Netherlands}
\affiliation{Department of Physics and Astronomy, VU University Amsterdam, 1081 HV Amsterdam, The Netherlands}

\author{Kees Steinebach}
\altaffiliation[Current address: ]{Department of Physics and Astronomy, VU University Amsterdam, 1081 HV Amsterdam, The Netherlands}
\affiliation{Van Swinderen Institute for Particle Physics and Gravity, University of Groningen, 9747 AG Groningen, The Netherlands}

\author{Rob G. E. Timmermans}
\affiliation{Van Swinderen Institute for Particle Physics and Gravity, University of Groningen, 9747 AG Groningen, The Netherlands}
\affiliation{Nikhef, National Institute for Subatomic Physics, 1098 XG Amsterdam, The Netherlands}

\author{Anno Touwen}
\affiliation{Van Swinderen Institute for Particle Physics and Gravity, University of Groningen, 9747 AG Groningen, The Netherlands}
\affiliation{Nikhef, National Institute for Subatomic Physics, 1098 XG Amsterdam, The Netherlands}

\author{Wim Ubachs}
\affiliation{Department of Physics and Astronomy, VU University Amsterdam, 1081 HV Amsterdam, The Netherlands}

\author{Lorenz Willmann}
\affiliation{Van Swinderen Institute for Particle Physics and Gravity, University of Groningen, 9747 AG Groningen, The Netherlands}
\affiliation{Nikhef, National Institute for Subatomic Physics, 1098 XG Amsterdam, The Netherlands}

\author{Yanning Yin}
\altaffiliation[Current address: ]{Department of Chemistry, University of Basel, 4056 Basel, Switzerland}
\affiliation{Van Swinderen Institute for Particle Physics and Gravity, University of Groningen, 9747 AG Groningen, The Netherlands}
\affiliation{Nikhef, National Institute for Subatomic Physics, 1098 XG Amsterdam, The Netherlands}

\begin{abstract}

A measurement of the magnitude of the electric dipole moment of the electron (eEDM) larger than that predicted by the Standard Model (SM) of particle physics is expected to have a huge impact on the search for physics beyond the SM. Polar diatomic molecules containing heavy elements experience enhanced sensitivity to parity ($P$) and time-reversal ($T$)-violating phenomena, such as the eEDM and the scalar-pseudoscalar (S-PS) interaction between the nucleons and the electrons, and are thus promising candidates for measurements. The NL-\textit{e}EDM collaboration is preparing an experiment to measure the eEDM and S-PS interaction in a slow beam of cold BaF molecules [Eur. Phys. J. D, 72, 197 (2018)]. Accurate knowledge of the electronic structure parameters, $W_d$ and $W_s$, connecting the eEDM and the S-PS interaction to the measurable energy shifts is crucial for the interpretation of these measurements.

In this work we use the finite field relativistic coupled cluster approach to calculate the $W_d$ and $W_s$ parameters in the ground state of the BaF molecule. Special attention was paid to providing a reliable theoretical uncertainty estimate based on investigations of the basis set, electron correlation, relativistic effects \textcolor{purple}{and geometry}. Our recommended values of the two parameters, including conservative uncertainty estimates, are 3.13 $\pm$ $0.12 \times 10^{24}\frac{\text{Hz}}{e\cdot \text{cm}}$ for $W_d$ and 8.29 $\pm$ 0.12 kHz for $W_s$.

\end{abstract}

\maketitle

%-------------------------
\section{Introduction}
%-------------------------

The Standard Model of particle physics (SM) is highly successful in providing predictions for laboratory experiments\cite{Tanabashi2018}. At the same time, it does not account for major cosmological observations, in particular the matter-antimatter asymmetry of the Universe\cite{DinKus03} and the existence of dark matter\cite{Ber10} and dark energy\cite{Peebles2003}. In order to address these, as well as other shortcomings of the SM, theoretical extensions have been proposed, such as Grand Unified Theories or supersymmetric models \cite{ArkHanMan16}. These proposed theories invariably predict new physical phenomena, such as the variation of fundamental constants in space and in time \cite{Uza11} or the violation of fundamental symmetries orders of magnitude larger than the SM predictions \cite{Commins1999,GinFla04}.% \rob{Btw for variation of contants you need something much more exotic than GUTs or SUSY ("standard" quantum field theory)...}

The search for the violation of fundamental symmetries provides unique opportunities to observe new phenomena and to test the various SM extensions. This search is conducted with different experimental methods, including precision measurements on atoms and molecules, which offer a low energy small-scale alternative\cite{GinFla04,SafBudDeM18} to the high-energy collider research in this field\cite{Rappoccio2019}. Such experiments take advantage of the many accessible energy levels in atoms and even more so in molecules due to their additional vibrational and rotational motions\cite{DeMille2015}. In addition, such systems can be manipulated and controlled with external fields in numerous ways. Atoms and molecules can be used to probe a wide variety of physical phenomena and in many cases experience strong enhancement effects that make the measurement of otherwise tiny signals possible.  

One of the most sought-after phenomena is that of the electric dipole moment of the electron (eEDM). A nonzero permanent eEDM would break the parity ($P$) and the time-reversal ($T$) symmetries and as a consequence of the $CPT$ theorem, assuming $CPT$ invariance, it also breaks the combined charge and parity ($CP$) symmetry. The \textcolor{blue}{$CP$} violation in the SM predicts an extremely small value of the eEDM ($\vert d_e \vert<10^{-38}$ $e$ $\cdot$ cm)\cite{PosRit05}, much too small to be measured \textcolor{blue}{with present day techniques}\cite{GavHerOrl94}. Many extensions of the SM, however, predict the eEDM to be orders of magnitude larger \cite{BerSuz91,Commins1999,CzaMar09}, and often within experimental reach. Thus, measurements of the eEDM (or constraints on its value) can provide a strong test of physics beyond the SM \cite{GinFla04, SafBudDeM18}.

It has been shown that the effect of an eEDM is strongly enhanced in paramagnetic polar diatomic molecules that contain heavy atoms\cite{Sandars1965, Sandars1967, Sushkov1978, Commins1999}. \textcolor{blue}{The presence of an unpaired electron possessing an eEDM induces a permanent molecular $P,T$-odd EDM which is many orders of magnitude larger than the eEDM itself. In fact, based on the predictions from SM extensions and the expected molecular enhancement factors, the resulting $P,T$-odd molecular EDM should be within experimental reach when using current technologies.} 

\textcolor{blue}{The above mentioned} enhancement \textcolor{blue}{of the eEDM} is due to relativistic contributions to the molecular structure in systems containing heavy nuclei\cite{Commins2007}. \textcolor{blue}{In the case of the eEDM,} it is described by the \textcolor{blue}{molecular enhancement factor} $W_d$, which is typically on the order of 1-100 GV/cm (expressed in units of electric field for comparison), whereas laboratory accessible electric fields are in the 1-10 kV/cm range. 

An additional \textcolor{blue}{contribution to the $P,T$-odd molecular EDM} is the scalar-pseudoscalar (S-PS) interaction between the electrons and the nucleons\cite{GinFla04, ChuRam15, Jung2013}. This effect is also enhanced in molecules containing heavy atoms; the corresponding \textcolor{blue}{molecular enhancement factor} is referred to as $W_s$. 

Furthermore, due to the presence of close-lying rotational levels of opposite parity, polar diatomic molecules can be easily polarized in laboratory accessible electric fields. \textcolor{blue}{As will be discussed further in Sec. \ref{sec:measurement}, a considerable amount of polarization} is crucial for a measurable effect. \textcolor{blue}{From the measured $P,T$-odd molecular EDM, the magnitude of the eEDM or S-PS interaction can be extracted using the enhancement factors $W_d$ and $W_s$, respectively.}

Several experiments aimed to take advantage of these enhancement effects and have used polar diatomic molecules to set upper limits on the eEDM\cite{Hudson2011,Baron2014,Cairncross2017, AndAngDem18}, with the current most stringent result, $|d_e| < 1.1 \times 10^{-29} ~ e \cdot \text{cm}$, achieved by the ACME collaboration using the excited $H^3 \Delta_1$ state of the ThO molecule\cite{AndAngDem18}.

In the recently established NL-\textit{e}EDM collaboration in Groningen and Amsterdam, The Netherlands, we are  setting up  an experiment to measure the eEDM and S-PS interaction using a slow beam of BaF molecules in the $X^2\Sigma_{1/2}^+$ ground state. The electronic structure of BaF \cite{HaoPasVis19} lends itself to efficient deceleration and laser-cooling, which should allow for creating an intense and slow beam of molecules with a sensitivity as low as of $5 \times 10^{-30} ~ e \cdot \text{cm}$.\cite{Aggarwal2018}
In this experiment, accurate knowledge of the $W_d$ and $W_s$ parameters, connecting the $P,T$-violating effects to the measurable energy shifts, \textcolor{blue}{see Sec. \ref{sec:measurement}}, is crucial for extracting the magnitude of these effects from the measurement.
\textcolor{blue}{As long as the magnitude of the eEDM and the S-PS interaction are not known,} $W_d$ and $W_s$ cannot be determined from experiment and since the precision of these parameters will ultimately impact the interpretation of the experiment, they should be provided from reliable high accuracy calculations. Such calculations should take into account both relativistic effects and electron correlation at the highest possible level. Furthermore, the nature of the computational method should be such as to allow us to set reliable uncertainties on our predictions.

The aim of this work is to provide these parameters for the ground $X^2\Sigma_{1/2}^+$ state of the BaF molecule at the highest level of theory currently available for heavy many-electron systems, namely using the single reference relativistic coupled cluster approach with singles, doubles, and perturbative triple excitations, CCSD(T). Furthermore, we perform a comprehensive investigation of the influence of various computational parameters on the results; this provides us with an in-depth insight into the effect of the $P,T-$violating phenomena on the molecular electronic structure. Even more importantly, it allows us to assign stringent error bars on our predictions. We also compare our results and the corresponding uncertainties to various previous studies\cite{Kozlov1995, Kozlov1997, Nayak2006, Nayak2007, Meyer2008, Fukuda2016, Gaul2017, Gaul2019, Abe2018, Sunaga2018, Talukdar2020} . 

The recommended values of the $W_d$ and the $W_s$ parameters, including the uncertainties, will be used in the interpretation of the NL-\textit{e}EDM experiment. 

%-------------------------
\section{Measurement interpretation and enhancement factors}\label{sec:measurement}
%-------------------------
The experimental strategy to search for a $P,T$-odd energy shift (due to the eEDM and S-PS interaction) in BaF molecules is as follows. \textcolor{purple}{Due to the nuclear spin, $I=1/2$, of $^{19}$F (the employed isotope of Ba is $^{138}$Ba which has $I=0$) each rotational level, labelled by $N$ in the case of a Hund's case b), is split into hyperfine levels labelled by $F$; with inclusion of the electron spin this yields two hyperfine levels with F=0 and F=1 in the lowest rotational level (N=0).} A superposition of two hyperfine substates ($m_F = \pm1$), which have their spin orientated in opposite directions in an applied magnetic field, is created in the ($N=0, F=1$) hyperfine level of the electronic, vibrational and rotational ground state of the molecule. The $N=0$ state is chosen partly due to its favourable polarization factor, which will be discussed at the end of this section. As the molecule flies through a region with carefully controlled electric and magnetic fields, $P,T$-odd interactions lead to a small energy shift between the components of the superposition which depend on the orientation of the electric field with respect to the magnetic field. This energy difference leads to a build-up of a phase difference which is subsequently read out when the superposition is projected onto the $N=0,F=0$ state. A $P,T$-odd signal is detectable through the determination of the difference in the accumulated phase for the parallel and the anti-parallel orientation of the magnetic and electric fields. The experimental resolution that can be reached in this signal depends on the fringe contrast, i.e. on the statistics, on the stability and homogeneity of the applied electric and magnetic fields, as well as on the velocity and flux stability of the molecular beam. It is attractive to use long coherent interaction times, as the sensitivity improves linearly with this parameter. The experimental approach is further detailed in Ref. \citenum{Aggarwal2018}.

Both the eEDM and the S-PS interaction contribute to the energy shift measured in the experiment. As we will see later, they are both electronic properties, which means that in the Born-Oppenheimer approximation the energy shift due to the associated operators is given by the expectation value over the electronic wave function, labeled by $|\eta,\Omega\rangle$. $\Omega$ is the projection of electronic angular momentum on the internuclear axis and $\eta$ represents the additional parameters specifying the \textcolor{purple}{vibronic} state. For simplicity, since we consider only one \textcolor{purple}{vibronic} state, in the following we will omit $\eta$.
The resulting energy shift is found to be proportional to $\Omega$.
For $\Omega\neq0$, the electronic states are two-fold degenerate and an additional $P,T$-odd term should be added to the spin-rotational Hamiltonian, $\hat{H}_{sr}$, which is a convenient way of describing the spin-rotational structure within a specific vibronic level. In the case of $|\Omega|=\frac{1}{2}$, the $P,T$-odd Hamiltonian takes the form\cite{Kozlov1995}:
\begin{equation}{\label{effectiveHamiltonian}}
    \hat{H}^{P,T}=(W_d d_e +W_s k_s)\vec{S}\cdot\hat{n} ,
\end{equation}
where $\vec{S}$ is the effective spin and $\hat{n}$ the internuclear axis. The effective spin is defined such that $\vec{S}\cdot\hat{n} |\Omega\rangle = \Omega |\Omega\rangle$.

With no external electric field, and neglecting parity violating effects for the moment, the molecular states are eigenstates of parity. It can be shown that these states are equal superpositions of states with $+\Omega$ and $-\Omega$ projections\cite{Brown2003}. Consequently, such a state will have zero $P,T$-odd energy shift, which can be seen from taking the expectation value of Eq. (\ref{effectiveHamiltonian}) over the rotational states, resulting in $\langle \vec{S}\cdot\hat{n}\rangle=0$. However, in an applied electric field, the zero-field states mix, giving rise to $\langle \vec{S}\cdot\hat{n}\rangle=P\Omega$, where $P$ is the polarization factor\cite{Tarbutt2009}.

\begin{figure}[t]
    \raggedleft
    \includegraphics[width=0.96\linewidth]{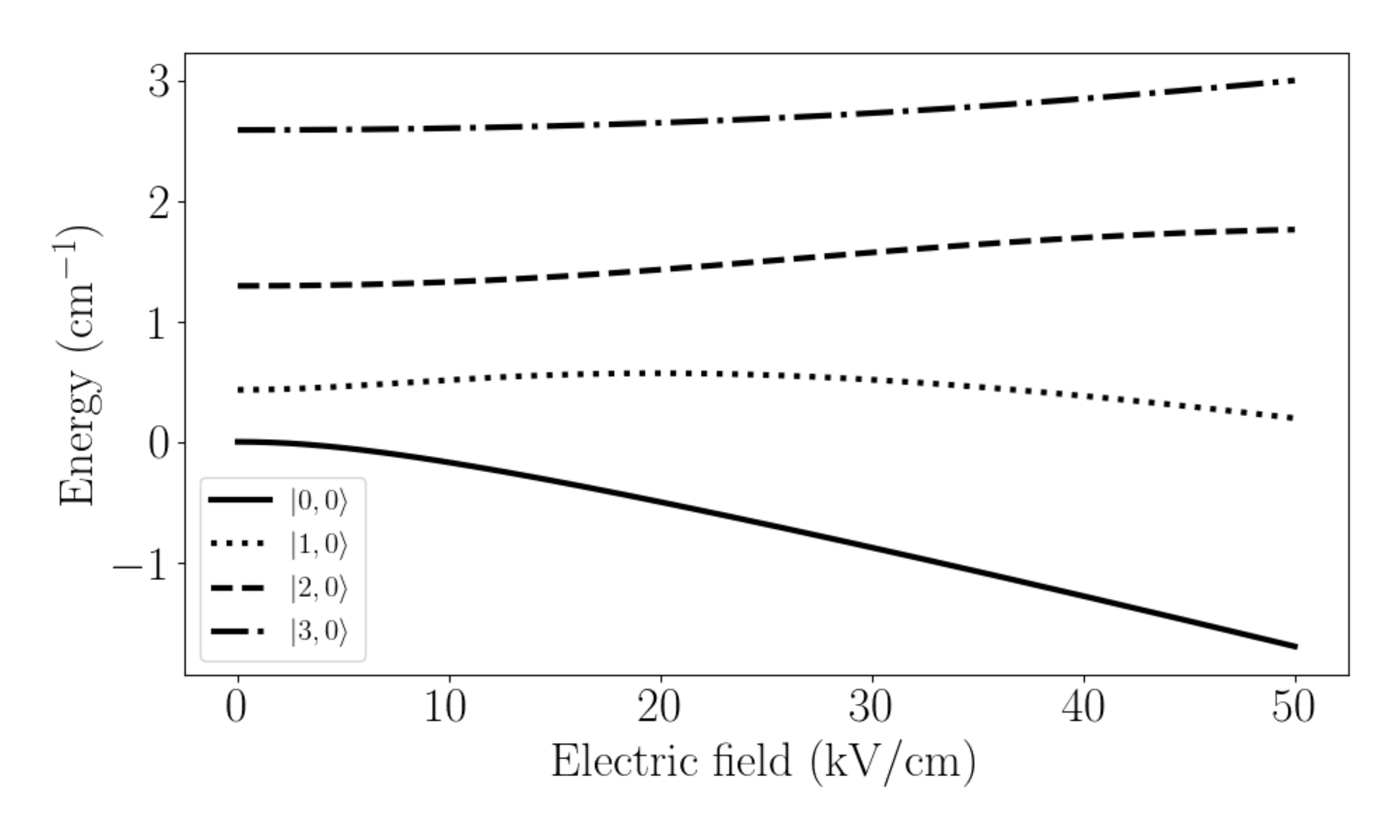}
    \includegraphics[width=\linewidth]{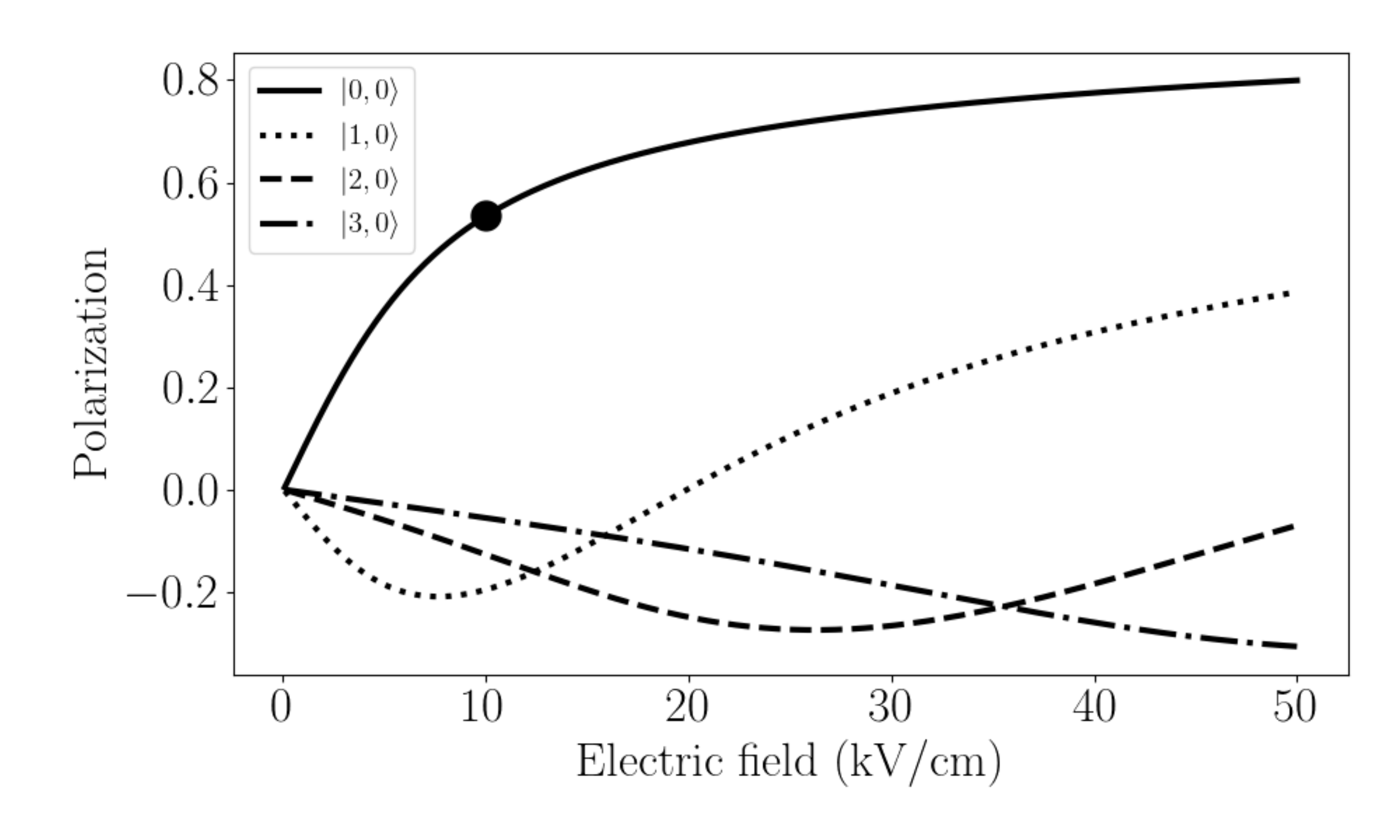}
    \caption{Top: Stark curves of the $M_N=0$ components of the lowest 4 rotational states, labelled by their zero-field states, in the $X ^2\Sigma_{1/2}^+$ state of BaF\textcolor{blue}{, $|N,M_N\rangle$}. Bottom: Corresponding polarization curves. The point shows the polarization at the typical experimental electric field strength of 10 kV/cm.}
    \label{fig:pol}
\end{figure}

A common way to estimate $P$ at a given field strength is to consider the Stark effect, since it similarly describes the mixing of opposite parity levels due to an external electric field. The polarization factor can then be related to the derivative of the Stark shift by\cite{Tarbutt2009,Lemeshko2013}:
\begin{eqnarray}\label{eq:pol}
P &=& -\frac{1}{d_{\text{mol}}} \frac{\partial \Delta W_{\text{Stark}}  }{\partial E_{\text{ext}}}, 
\end{eqnarray}
where $\Delta W_{\text{Stark}}$ is the energy shift due to the Stark Hamiltonian $\hat{H}_{\text{Stark}} = -\vec{d}_{\text{mol}}\cdot\vec{E}_{\text{ext}}$ and $\vec{d}_{\text{mol}}$ is the usual molecular dipole moment directed along the internuclear axis. \textcolor{blue}{$\vec{d}_{\text{mol}}$ differs from the aforementioned $P,T$-odd molecular EDM in the sense that the derivative of the associated Stark shift, $\Delta W_{\text{Stark}}$, goes to zero when $E_{\text{ext}}=0$. This is not the case for the $P,T$-odd molecular EDM which is consequently referred to as a permanent EDM.} 

Due to the $P,T$-odd energy shift the $P,T$-odd phase accumulated during the coherent interaction time $\tau$ can be expressed as:
\begin{eqnarray}
\phi^{P,T} = (W_d d_e + W_s k_s) \Omega |P| \tau .
\end{eqnarray}

Note that a measurement on single molecular species provides only a linear combination of the two effects; in order to disentangle these, measurements on multiple systems with different sensitivities (or different $W_d$/$W_s$ ratios) are needed \cite{Jung2013, FleJun18}. Two recent studies discuss possible optimum combinations of molecules for measurements \cite{SunAbePra19,Gaul2019}.

In order to illustrate the importance of the polarization factor, we have calculated the Stark shift of the $X ^2\Sigma_{1/2}^+$ state in BaF by evaluating the sum of the effective rotational Hamiltonian and the Stark Hamiltonian: 
\begin{eqnarray}
\hat{H} = B \vec{N}^2 - {d}_{\text{mol}} \hat{n} \cdot \vec{E}_{\text{ext}} ,
\end{eqnarray}
where $\vec{N}$ is the rotational \textcolor{blue}{angular momentum of the nuclei} with projection $M_N$ on a space fixed axis, in the basis of the 10 lowest rotational levels and diagonalizing the resulting 66$\times$66 matrix. Note that for simplicity, spin-rotation and hyperfine terms have been neglected as these would have a minor impact on the plotted curves. In the top panel of Fig. \ref{fig:pol} we show the Stark shift of the $M_N=0$ component of the 4 lowest rotational levels, labelled by their zero-field state, as a function of the external electric field. In the bottom panel of Fig. \ref{fig:pol} the corresponding polarization factor is shown, calculated using Eq. (\ref{eq:pol}), revealing very different behavior of the polarization effect for the various rotational levels, which has a decisive influence on the resulting precision of an eEDM experiment. %This is one of the 

For the field strengths shown in Fig. \ref{fig:pol}, the $N=0$ level provides by far the largest polarization factor, which is one of the reasons for using this level in the ongoing eEDM measurements\cite{Aggarwal2018,Hudson2011}. In a typical experiment a field strength of $\sim$ 10 kV/cm can be achieved, leading to a polarization factor of $\sim$ 50\% in the $N=0$ level, as indicated by the point in the bottom panel of Fig. \ref{fig:pol}. 
%\pabh{How accurately can we know this factor? The point on the figure is at 0.537, should we use this value? On the other hand, the precise value depends on the actual electric field used, which I guess we don't know very precisely yet.}

The remainder of this article will be concerned with the calculations of the electronic enhancement factors, $W_d$ and $W_s$.

%-----------------
\section{Theory}
%-----------------

In Eq. (\ref{effectiveHamiltonian}) the EDM interaction energy was described using an effective spin Hamiltonian. In the following, we will present the overall framework that allows us to calculate $W_d$ and $W_s$ with \textit{ab initio} methods. Throughout this section atomic units will be used. As a starting point, one needs to consider the eEDM and S-PS Lagrangian densities\cite{Salpeter1958, Commins1999}. In the case of the eEDM interaction, the Lagrangian density, written in Dirac notation, leads to the following Hamiltonian:
\begin{eqnarray}\label{eq:EandB}
\hat{H}^{\text{eEDM}} = -d_e ({\gamma}^0 \vec{{\Sigma}}\cdot\vec{E} + i\vec{{\gamma}}\cdot\vec{B} ),
\end{eqnarray}
where ${\gamma}^0 $ and $\vec{{\gamma}}$ are the usual Dirac gamma matrices, $\vec{{\Sigma}} = \begin{pmatrix} \vec{{\sigma}} & 0 \\ 0 & \vec{{\sigma}}
\end{pmatrix}$ is the vector of Pauli spin matrices:
\begin{equation}\label{eq:pauli}
{\sigma}_x = 
\begin{pmatrix}
0 & 1 \\
1 & 0 
\end{pmatrix}
, ~~
{\sigma}_y = 
\begin{pmatrix}
0 & -i \\
i & 0 
\end{pmatrix}
, ~~
{\sigma}_z = 
\begin{pmatrix}
1 & 0 \\
0 & -1 
\end{pmatrix},
\end{equation} 
and $\vec{E}$ and $\vec{B}$ are the total electric and magnetic fields. The contribution from the interaction of $d_e$ with the magnetic field has been shown to be small and thus the second term is usually omitted\cite{Lindroth1989}. 

The electric field in Eq.(\ref{eq:EandB}) includes both the external and the internal electric fields, the latter due to the atomic nuclei and the other electrons in the system. Consequently, $\vec{E}$ is a two-body operator and it is advantageous to rewrite the first term of Eq. (\ref{eq:EandB}) into the one-body operator form\cite{Martensson-Pendrill1987}:
\begin{equation}\label{eq:eEDM}
\hat{H}^{\text{eEDM}} = 2 i c d_e \sum_i {\gamma}^5_i {\gamma}^0_i \vec{p}_i^{~2},
\end{equation}
where ${\gamma}^5=i{\gamma}^0{\gamma}^1{\gamma}^2{\gamma}^3$ and $\vec{p}_i$ is the momentum of electron $i$. \textcolor{blue}{Note that the eEDM operator in Eq. (\ref{eq:eEDM}) is the \textcolor{purple}{dominating} one in the expression for the eEDM Hamiltonian, \textcolor{purple}{based on numerical calculations on Cs}, derived from a more advanced, and correct to the order $\alpha^2$, expression based on the inclusion of the full Breit two-electronic interaction in addition to the Coulomb one (see the expression (3.4) in Ref. \citenum{Lindroth1989}).}

The S-PS interaction can be described by the following Hamiltonian\cite{Ginges2004}:
\begin{eqnarray}\label{eq:ws}
\hat{H}^{\text{S-PS}} = i \frac{G_F}{\sqrt{2}} Z_N k_s \sum_i {\gamma}^0_i {\gamma}^5_i \rho(\vec{r}_{iN}),
\end{eqnarray}
where $G_F$ is the Fermi constant (2.2225$\times10^{-14}$ in atomic units), $Z_N$ is the atomic number of nucleus $N$ and $\rho(\vec{r}_{iN})$ is the nuclear charge distribution in the form of a Gaussian distribution. In principle Eq. (\ref{eq:ws}) should include the sum over all nuclei in the system. We however consider here only the contribution from the Ba nucleus, since the contribution from the F atom will be negligible due to two factors; 1) the molecular orbital of the unpaired electron is of mainly Ba 6s character, which has a much larger overlap with the Ba nucleus than with the F one and 2) more importantly, both the eEDM and the S-PS interaction scale as $Z^3$, which greatly suppresses the F contribution\cite{GinFla04}. Note that sometimes the S-PS operator is parametrized in terms of $C_s$ instead of $k_s$; the two parameters are related by $Z k_s = A C_s$, where $A$ is the mass number\cite{Commins1999}. 

For variational wave functions the $W_d$ and $W_s$ parameters can be evaluated as the expectation values of the eEDM or S-PS operators (Eqs. (\ref{eq:eEDM}) and (\ref{eq:ws})). For non-variational wave functions, such as the coupled cluster (CC) wave function, the finite field method is a convenient way to evaluate molecular properties and we use it to calculate both $W_d$ and $W_s$. 
The advantage of this formulation is in the simplicity of its implementation and in the fact that no truncation of the CC expansion is necessary. We have recently applied this implementation to calculations of the anapole-moment enhancement parameter $W_A$ and the hyperfine structure (HFS) parameters in BaF \cite{Hao2018, Haase2020a}; for the hyperfine-structure constants, where experimental values are available, we have achieved accuracy on a single percent level, providing a confirmation of the reliability of this approach.

In the finite field method the Hamiltonian describing the perturbation, $\hat{H}_k$, is added to the unperturbed Hamiltonian, $\hat{H}^{(0)}$, with a pre-factor (sometimes denoted the field strength), $\lambda$\cite{Cohen1965,Visscher1998}:
\begin{equation} \label{eq:H_tot}
\hat{H} = \hat{H}^{(0)} + \lambda_k \hat{H}_k,
\end{equation}
where $\hat{H}_k$ is the eEDM operator, $\hat{H}^{\text{eEDM}}/d_e$, or the S-PS operator, $\hat{H}^{\text{S-PS}}/k_s$ (Eqs. (\ref{eq:eEDM}) and (\ref{eq:ws})) and $\lambda_k$ represents the effective $d_e$ or $k_s$, respectively. In our case, $\hat{H}^{(0)}$ is the relativistic Dirac-Coulomb Hamiltonian:
\begin{equation}\label{eq:DC}
\hat{H}^{(0)} = \sum_i \left[
{\beta}_i m c^2
+ c \vec{{\alpha}}_i \cdot \hat{\vec{p}}_i
- V_{\text{nuc}}(r_i)
\right]
+ \frac{1}{2} \sum_{i\ne j} \frac{1}{r_{ij}},
\end{equation}
where $\vec{{\alpha}}$ and ${\beta}$ are the Dirac matrices:
\begin{equation}
\vec{{\alpha}} = 
\begin{pmatrix}
0 & \vec{{\sigma}} \\
\vec{{\sigma}} & 0 \\
\end{pmatrix}
, ~~
{\beta} = \begin{pmatrix}
1_{2\times2} & 0 \\
0 & -1_{2\times2}
\end{pmatrix}
\end{equation}
and $\vec{{\sigma}}$ is the vector consisting of the Pauli spin matrices, defined in Eq. (\ref{eq:pauli}).

As a consequence of the introduction of $\lambda_k$ in Eq. (\ref{eq:H_tot}), the energy can be expanded in a Taylor series around $\lambda_k=0$:
\begin{equation}
E_{\Omega}(\lambda_k) = E_{\Omega}^{(0)} + \lambda_k E^{(1)}_{\Omega} + \mathcal{O}(\lambda_k^n),
\end{equation}
where $\mathcal{O}(\lambda_k^n)$ denotes higher-order terms. $E_{\Omega}$ is the total energy of a given electronic state in the presence of the perturbation $\hat{H}_k$. $E_{\Omega}^{(1)}$ is proportional to the eEDM or S-PS energy shift presented in Eq. (\ref{effectiveHamiltonian}) with the proportionality factors being $d_e$ or $k_s$, respectively. The magnitude of $\lambda_k$ can be chosen such that higher order terms vanish, and $W_d$ and $W_s$ (below represented by $W_k$) can be obtained as the first derivative of the energy with respect to $\lambda_k$:
\begin{equation} \label{eq:Wd_derivative}
W_k = \frac{1}{\Omega} \left. \frac{d E_{\Omega}(\lambda_k)}{d \lambda_k} \right| _{\lambda_k=0}.
\end{equation}

In practice, $E_\Omega(\lambda_k)$ is calculated at different values of $\lambda_k$ and $W_k$ are obtained by numerical differentiation. Due to the eEDM and S-PS operators being $T$-odd, we evaluate them at the Kramer's unrestricted CC level, which uses unperturbed Kramer's restricted Hartree-Fock orbitals\cite{Visscher1996}.

It should be mentioned that $W_d$ is often referred to in literature in terms of the so-called effective electric field, $\eeffm$; the relation between the two is given by $W_{d} = \eeffm / \Omega$.

%---------------------------------
\section{Computational details}
%---------------------------------

All the calculations were carried out with a modified version of the DIRAC17 program\cite{Dirac17,Saue2020}. If not stated otherwise, the default program settings were used. We implemented the eEDM operator, Eq. (\ref{eq:eEDM}), at the unrestricted CC level whereas the S-PS operator was constructed from the existing repository of one-electron operators. 

Dyall's \textcolor{blue}{uncontracted} all-electron relativistic basis sets of double-, triple- and quadruple-zeta quality (denoted vdz, vtz and vqz) were employed\cite{Dyall2009,Dyall2016}, both in their original form as well as with additional tight and diffuse functions added in an even-tempered fashion as described in Section \ref{sec:results}. In addition, also the cvXz and aeXz basis sets were used, which include additional functions with large exponents (so-called tight functions) needed to correlate the core-valence region and all the electrons, respectively. The singly augmented vXz basis sets (s-aug-vXz), which include a set of automatically generated diffuse functions for each symmetry block, were also used. These functions are generated in an even-tempered fashion and are shown in the Supplementary Information Tab. S1. \textcolor{blue}{The Dyall basis sets provide exponents for the large component basis functions and the small component basis functions were constructed using the restricted kinetic balance.} In the calculation of the Fock matrix, a tight threshold on the screening of two-electron integrals of $10^{-15}$ a.u. was used.

The unrestricted CC method with single and double excitations was employed along with three schemes for the inclusion of perturbative triple excitations: +T, (T) and $-$T\cite{Visscher1996}. The difference between these schemes is the number of triple excitations that are included; in CCSD+T, these are all the excitations up to 4th order in perturbation theory, in CCSD(T) a subset of 5th order excitations is also included, and in CCSD$-$T, one additional 5th order excitation is added. In addition, also the Fock-space coupled cluster approach  (FSCC) was used\cite{Visscher2001}. If not stated otherwise, all electrons were included in the correlation calculation with a cut-off of virtual orbitals set at 2000 a.u. 

\textcolor{blue}{The convergence threshold was $2\times10^{-10}$ a.u. on the DHF energy and $10^{-12}$ a.u. on the CC amplitudes. }
The field strengths employed in the finite field procedure were $0.0$ and $\pm10^{-9}$\textcolor{blue}{, resulting in an energy change on the order of $10^{-9}$ a.u. At this field strength, the perturbed energy behaves linearly with respect to the field strength and }the numerical differentiation was obtained through linear regression analysis \textcolor{blue}{through the three points. Based on an examination of the field strength dependence we expect the uncertainty due to numerical precision to be not larger than 0.1\%.}

The experimental BaF bond length of 2.162 \AA~was taken from the NIST Chemistry WebBook \cite{NISTBaF,Knight1971}.

%-----------------------------------
\section{Results and discussion}\label{sec:results}
%-----------------------------------

%-----------------------------------
\subsection{Basis set}\label{sec:basis}
%-----------------------------------

%\begin{table}[t]
%\begin{tabular}{lcc}
%\hline
%basis set & \eeff [GV/cm] & Ws [kHz] \\
%\hline
%%vdz & 6.342 & 7.552 \\
%%vtz & 6.449 & 8.174 \\
%%vqz & 6.398 & 8.184 \\
%%s-aug-vqz & 6.393 & 8.153 \\
%%cvqz & 6.477 & 8.29? \\
%vdz       & 6.34 & 7.55 \\
%vtz       & 6.45 & 8.17 \\
%vqz       & 6.40 & 8.18 \\
%cvdz      & 6.42 & 7.64 \\
%cvtz      & 6.55 & 8.30 \\
%cvqz      & 6.48 & 8.29 \\
%aeqz      & 6.47 & 8.31 \\
%s-aug-vdz & 6.23 & 7.43 \\
%s-aug-vtz & 6.40 & 8.10 \\
%s-aug-vqz & 6.39 & 8.15 \\
%\hline
%\end{tabular}
%\caption{\protect\pabh{Should be written} Basis set. CCSD(T)} \label{tab:basis}
%\end{table}

\begin{table}[b]
\caption{Dependence of the calculated $W_d$ and $W_s$ on the quality of the basis set within the valence (v), core-valence (cv), and singly augmented valence (s-aug-v) families. The calculations were carried out on the CCSD(T) level.} \label{tab:basis}
\begin{tabular}{lccccccc}
\hline
 & \multicolumn{3}{c}{$W_d$ [$10^{24}\frac{\text{Hz}}{e\cdot \text{cm}}$]} & & \multicolumn{3}{c}{$W_s$ [kHz]} \\
\hline
X & vX & cvX & s-aug-vX & & vX & cvX & s-aug-vX  \\
\cline{2-4}\cline{6-8}
dz & 3.067  & 3.103  & 3.013  &  & 7.552 & 7.637 & 7.426 \\
tz & 3.119  & 3.166  & 3.096  &  & 8.174 & 8.300 & 8.100 \\
qz & 3.094  & 3.132  & 3.092  &  & 8.184 & 8.290 & 8.174 \\
\hline
\end{tabular}
\end{table}

We start our computational study with an analysis of the basis set dependence of the $W_d$ and $W_s$ factors. This allows us to select a basis set which is sufficiently converged for these properties and at the same time to estimate the error that is introduced by truncating the basis set at the chosen point. Since the standard basis sets are optimized for describing properties related to the valence region, such as bonding, and not core-dependent properties like $W_d$ and $W_s$, such a study becomes indispensable. To capture the full effect of varying the basis sets, the CCSD(T) method is used in this investigation. 

\begin{figure}[t!]
\includegraphics[width=\linewidth]{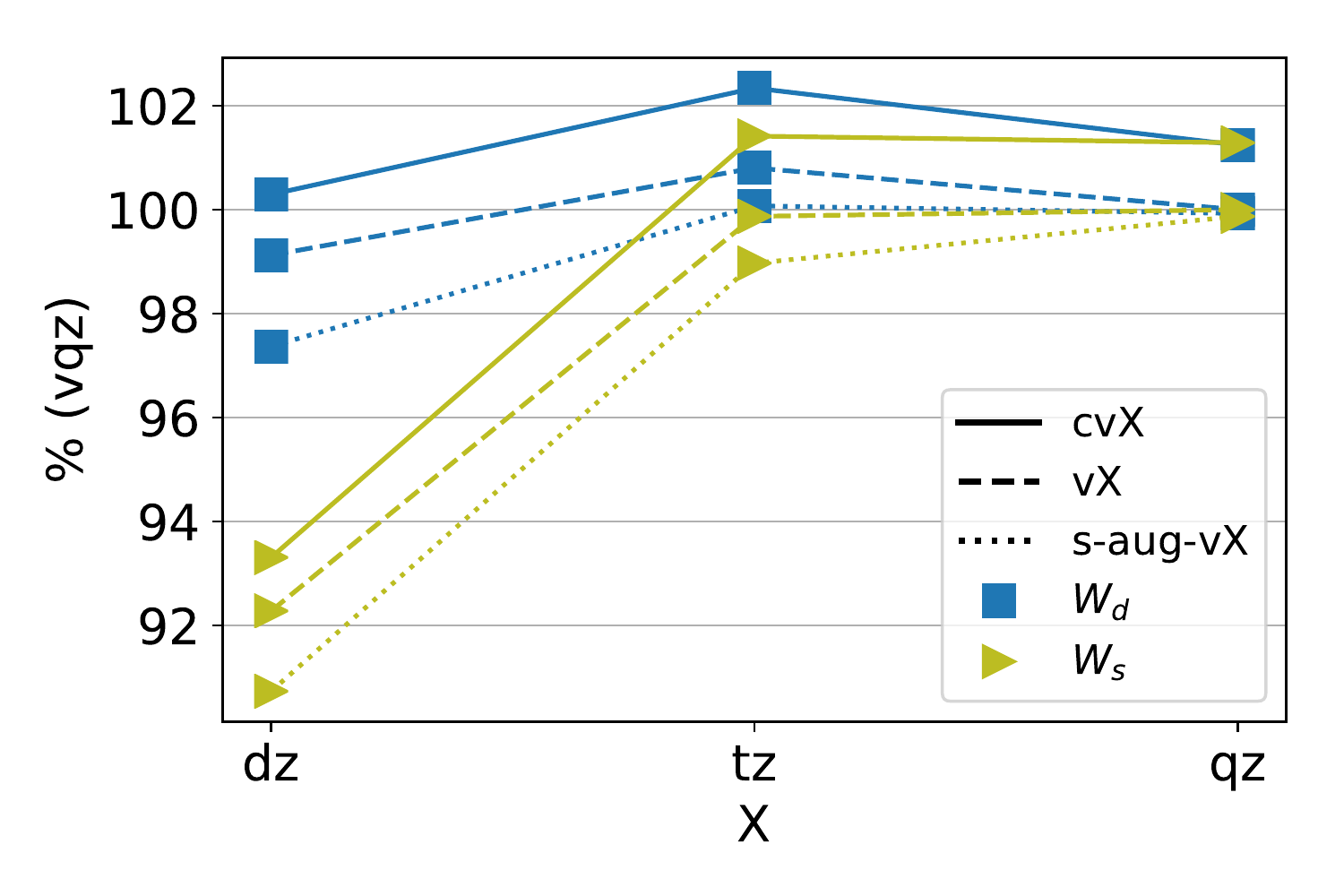}
\caption{Dependence of the calculated $W_d$ and $W_s$ on the basis set quality and family. The results are presented in percent with respect to the vqz values. The calculations were carried out on the CCSD(T) level (see Table \ref{tab:basis} for the values).} \label{fig:basis}
\end{figure}

In Table \ref{tab:basis} and Figure \ref{fig:basis}, $W_d$ and $W_s$ are shown at increasing basis set quality, i.e. cardinal number, and using three different families, namely the valence (vXz), core-valence (cvXz) and singly augmented valence (s-aug-vXz) families. The first thing to notice is that the two properties show different dependencies on the basis set quality. Going from double- to quadruple-zeta quality, $W_s$ generally shows a convergent behavior, whereas $W_d$ increases from double- to triple-zeta and decreases going from triple- to quadruple-zeta. This zig-zag behavior can be attributed to the competing effects of tight functions which increase $W_d$ and $W_s$ (demonstrated by comparison of cvXz to vXz) and diffuse functions which decrease $W_d$ and $W_s$ (s-aug-vXz compared to vXz). As the standard basis sets are constructed to optimise the description of the valence region, one can expect mostly diffuse functions to be added when increasing the basis set cardinal number. Consequently, the present results indicate that $W_s$ is more sensitive to the description of the valence region than $W_d$.

\begin{figure}[t]
\includegraphics[width=\linewidth]{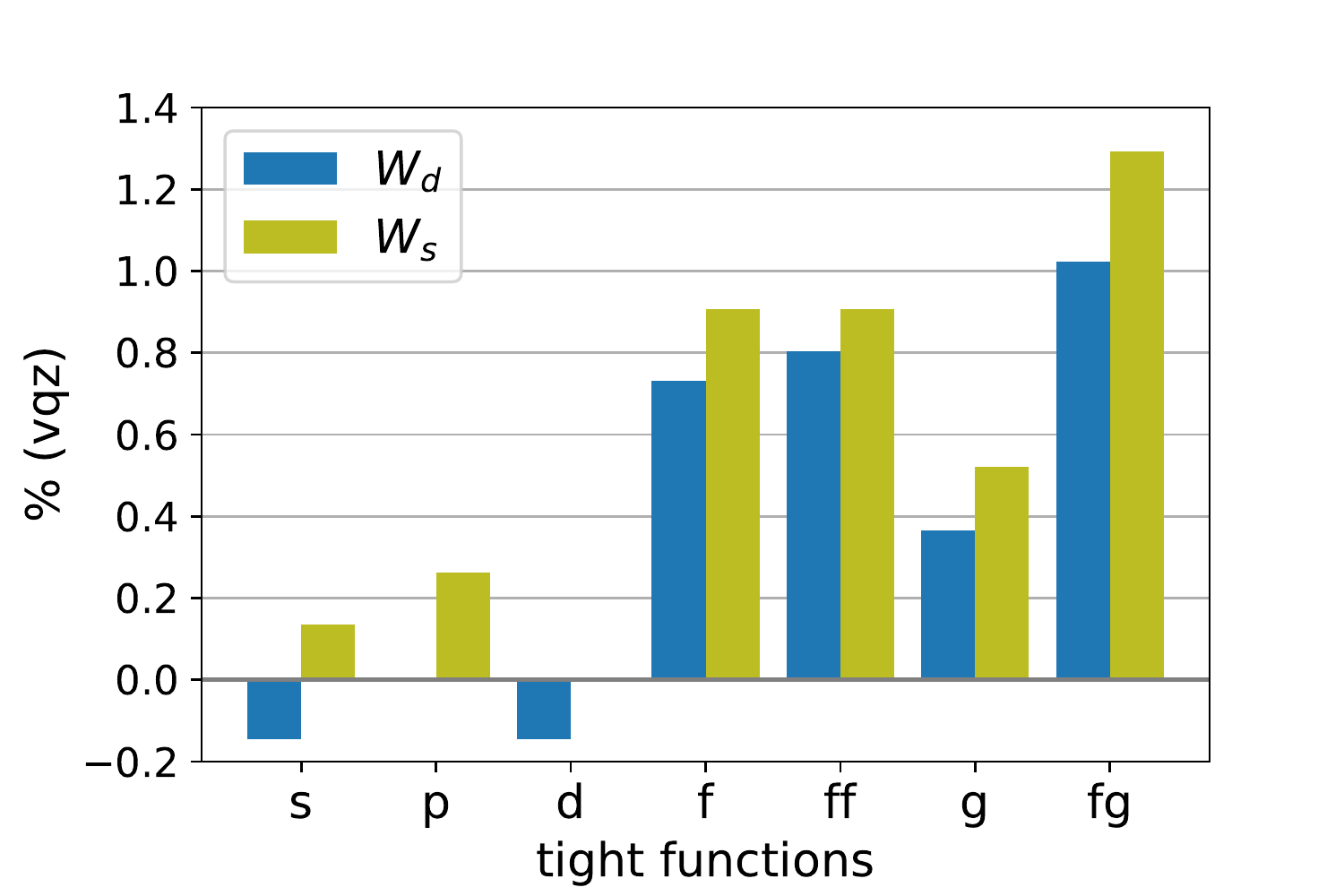}
\caption{Effect of additional tight (large exponent) functions of different symmetries on the calculated $W_d$ and $W_s$. The results are presented in percent with respect to the vqz values. The calculations were carried out on the CCSD(T) level.} \label{fig:tight}
\end{figure}

At the quadruple-zeta level, the effect of correlating the core-valence region (using the cvqz basis set) is very similar, i.e. $\sim$1.2\% for $W_d$ and $\sim$1.3\% for $W_s$, compared to the vqz basis. The difference between the vqz and cvqz basis sets lies in the addition of tight functions (i.e. functions with large exponents) with high angular momentum needed to correlate the 4d shell, that is 3 f-, 2 g- and 1 h- functions. To investigate which of these functions are responsible for the observed change in the $W_s$ and $W_d$ parameters, and the effect of additional tight functions of lower angular momentum, we performed an additional study, where we added tight functions of different symmetries manually, one by one. The results are shown in Fig. \ref{fig:tight} and the data points as well as the exponents which were generated in an even-tempered fashion can be found in Tables S2 and S3 in Supplementary Information. We can conclude that the vqz basis set is sufficiently saturated with respect to tight s-, p- and d-functions; adding extra tight functions in these symmetries has a small effect on the results (that we account for in the uncertainty estimation). Furthermore, the bulk of the effect is seen when adding a single tight f-function; the addition of a second tight f-function, however, leaves the calculated enhancement factors virtually unchanged. The effect of adding a tight g-function is around half of the f-function effect and the two contributions turn out to be roughly additive. This behaviour is very similar for the two parameters. 

To investigate whether additional correlation functions for the core-electrons have an effect on the two parameters we performed two calculations using the all-electron basis set at the quadruple-zeta level (aeqz), the results being $W_d=3.128 \times 10^{24}\frac{\text{Hz}}{e\cdot \text{cm}}$ and $W_s = 8.311$ kHz, changing the cvqz values by  $<$ 0.25\%; we will account for this effect in the uncertainty estimation.
%
%The effect of adding core-valence correlating functions here is slightly lower than was observed for HFS constants where it was $\sim$ 2 \%\cite{Haase2020a}. %Whereas both tight f- and g-functions contribute for $W_d$ and $W_s$, only tight f-functions made a significant contribution to the hfs constants. \pabh{Why this difference?}

\subsection{Correlation treatment}

In the following, we investigate the effect of various parameters related to electron correlation. In coupled cluster calculations of valence properties, such as polarizabilities\cite{Lim2004}, it is often sufficient to correlate only part of the electrons by freezing a number of core shells. Furthermore, only correlating the outer shells allows one to reduce the number of required virtual orbitals, thus significantly lowering the computational costs. However, in properties which depend on the description of the core region, such as $W_d$ and $W_s$ (or the less exotic hyperfine structure constants \cite{Haase2020a}) it is important to also correlate the core electrons and consequently to include a large number of virtual orbitals\cite{Talukdar2018a, Talukdar2020, Skripnikov2015, Skripnikov2017}. To facilitate a more quantitative insight into this behaviour, we investigate in the following how $W_d$ and $W_s$ depend on the size of the occupied and the virtual correlation space. To reduce the computational cost, the aetz basis set was used.

In the following two subsections the results were obtained with the CCSD(T) method. Subsection \ref{sec:corr} contains a comparison of the performance of the various schemes for inclusion of the perturbative triple excitations.

%-----------------------------------
\subsubsection{Active occupied space} \label{sec:active_elec}
%-----------------------------------

\begin{figure}[t!]
\includegraphics[width=\linewidth]{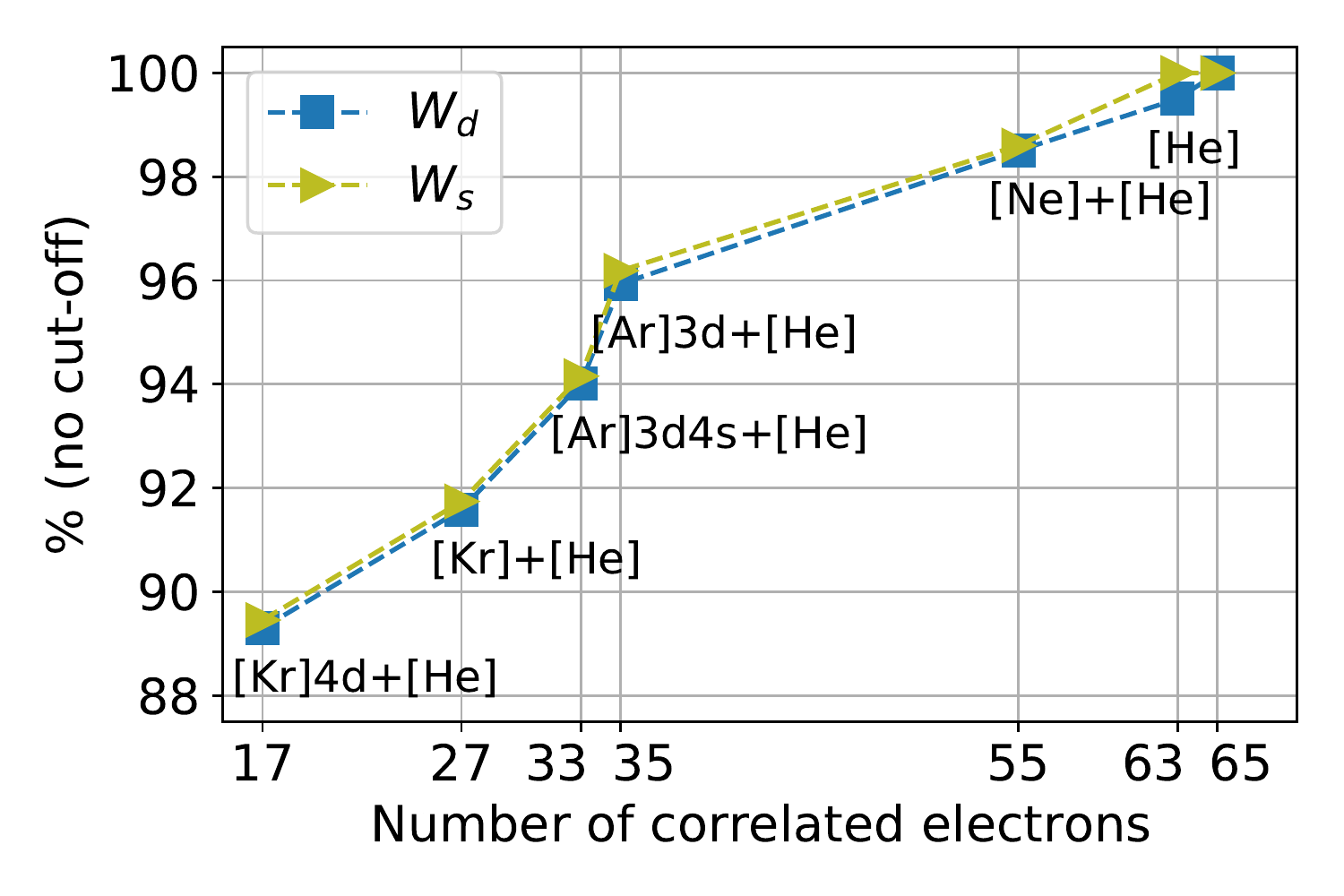}
\caption{Effect of correlating different number of electrons on the calculated $W_d$ and $W_s$, relative to an all-electron calculation. The calculations were performed on CCSD(T)/aetz level with a virtual cut-off of 2000 a.u. See also Table \ref{tab:active_electrons}. \textcolor{blue}{For each point, the frozen orbitals are shown (Ba+F).}} \label{fig:act_elec_vdz_vtz}
\end{figure}

Here we investigate the effect of excluding  a given number of electrons from the correlation treatment. In this study, regardless of the number of correlated electrons, all virtual orbitals below a cut-off of 2000 a.u. were included in the calculation as it has been shown that high-lying virtual orbitals are important for capturing all the correlation effects related to the core electrons\cite{Talukdar2018a, Talukdar2020, Skripnikov2015, Skripnikov2017}. 

In Fig. \ref{fig:act_elec_vdz_vtz}, the behavior of $W_d$ and $W_s$ is shown when increasing the number of correlated electrons from 17 to the entire  65; the results are shown in percent, with respect to those obtained in the all-electron calculations. We observe very similar behaviour for the two constants. When correlating only the 17 outer-core-valence electrons both properties are underestimated by around 10\%. In Ref. \citenum{Skripnikov2015} this effect was estimated to be about 5\% for $W_d$ in ThO and in Ref. \citenum{Skripnikov2017} around 4\% for $R_s$ (atomic analogue of $W_s$) in the Fr atom, which indicates that the importance of both core and core-valence correlation \textcolor{blue}{might have been} underestimated in the past\textcolor{blue}{, keeping in mind that the correlation effects could be different in the mentioned systems, for example due to a different electronic state.}

\begin{table}[t]
\renewcommand*{\arraystretch}{1.25}
\caption{Effect of the number of correlated electrons on the calculated $W_d$ and $W_s$. \textcolor{blue}{The frozen} orbitals are also shown, along with the relative deviation in percent compared to correlating all 65 electrons. The calculations were carried out on the CCSD(T)/aetz level.} \label{tab:active_electrons}
\begin{tabular}{cllcccc}
\hline
nr. \textcolor{blue}{corr.}      & \multicolumn{2}{c}{frozen orbitals} & $W_d$ & \% (65) & W$_s$ & \% (65) \\
elec.     & Ba & F & [$10^{24}\frac{\text{Hz}}{e\cdot \text{cm}}$] & & [kHz] & \\
\hline
17\footnote{outer-core-valence orbitals: Ba 6s, 5p, 5s and F 2p, 2s }
    & \textcolor{blue}{[Kr]4d  }   &         &  2.832 & 89.3 & 7.416 & 89.5  \\
27  & \textcolor{blue}{[Kr]    }   &         &  2.904 & 91.6 & 7.605 & 91.7  \\
33  & \textcolor{blue}{[Ar]3d4s}   &         &  2.981 & 94.0 & 7.805 & 94.2  \\
35  & \textcolor{blue}{[Ar]3d  }   &         &  3.042 & 95.9 & 7.974 & 96.2  \\
55  & \textcolor{blue}{[Ne]    }   & \textcolor{blue}{[He]}    &  3.123 & 98.5 & 8.174 & 98.6  \\
63  & \textcolor{blue}{[He]    }   &         &  3.155 & 99.5 & 8.29 & 100    \\
65  & --         & --      &  3.171 & 100  & 8.29 & 100    \\
\hline
\end{tabular}
\end{table}
%\begin{table}[b]
%\renewcommand*{\arraystretch}{1.25}
%\caption{Effect of the number of correlated electrons on the calculated $W_d$ and $W_s$. The largest atomic orbital contributions to the molecular orbitals frozen in each energy interval are also shown, along with the relative deviation in percent compared to correlating all 65 electrons. The calculations were carried out on the CCSD(T)/aetz level.} \label{tab:active_electrons}
%\begin{tabular}{cllcccc}
%\hline
%nr.       & \multicolumn{2}{c}{frozen orbitals} & $W_d$ & \% (65) & W$_s$ & \% (65) \\
%elec.     & Ba & F & [$10^{24}\frac{\text{Hz}}{e\cdot \text{cm}}$] & & [kHz] & \\
%\hline
%17\footnote{outuer-core-valence orbitals: Ba 6s, 5p, 5s and F 2p, 2s }
%          &   1s2p2s3d3p3s4s4p4d         &         &  2.832 & 89.3 & 7.416 & 89.5  \\
%27        &   1s2p2s3d3p3s4s4p         &         &  2.904 & 91.6 & 7.605 & 91.7  \\
%33        &   1s2p2s3d3p3s4s         &         &  2.981 & 94.0 & 7.805 & 94.2  \\
%35        &   1s2p2s3d3p3s &         &  3.042 & 95.9 & 7.974 & 96.2  \\
%55        &   1s2p2s     & 1s      &  3.123 & 98.5 & 8.174 & 98.6  \\
%63        &   1s         &         &  3.155 & 99.5 & 8.29 & 100    \\
%65        &   --         & --      &  3.171 & 100  & 8.29 & 100    \\
%\hline
%\end{tabular}
%\end{table}

Including 35 electrons in the correlation treatment, which corresponds to the often used cut-off of $-$20 a.u., has a noticeable effect and reduces the deviation to $\sim$ 4\%. A similar effect was found in Ref. \citenum{Talukdar2020}. This corresponds to including the 4s, 4p, and 4d electrons of Ba in the correlation treatment, as can be seen in Table \ref{tab:active_electrons}, where the main atomic orbital contributions (determined from a Mulliken population analysis) to the molecular orbitals frozen in each interval are listed. 

Consequently, excluding what can be considered as the core electrons, namely 1s - 3d on Ba and 1s on F, in total 30 electrons, has a considerable effect ($\sim$ 4\%) if one pursues high accuracy and low uncertainty. Even freezing the 1s, 2s and 2p electrons on Ba has a non-negligible effect of $\sim$ 1.5 \%. Conversely, this effect was found to be negligible in the case of $W_d$ and $W_s$ in YbF in Ref. \citenum{Sunaga2016}, which could be due to the absence of high-lying virtual orbitals from the calculation (due to a virtual space cut-off of 200 a.u.). In the same study, the effect of including the 3rd shell of Yb was found to be 2\% - 3\%, which agrees with our findings.

All electrons were correlated in the results which will be presented in the following and in the final recommended values. 

%-----------------------------------
\subsubsection{Active virtual space}
%-----------------------------------

\begin{figure}[t!]
\centering
\includegraphics[width=\linewidth]{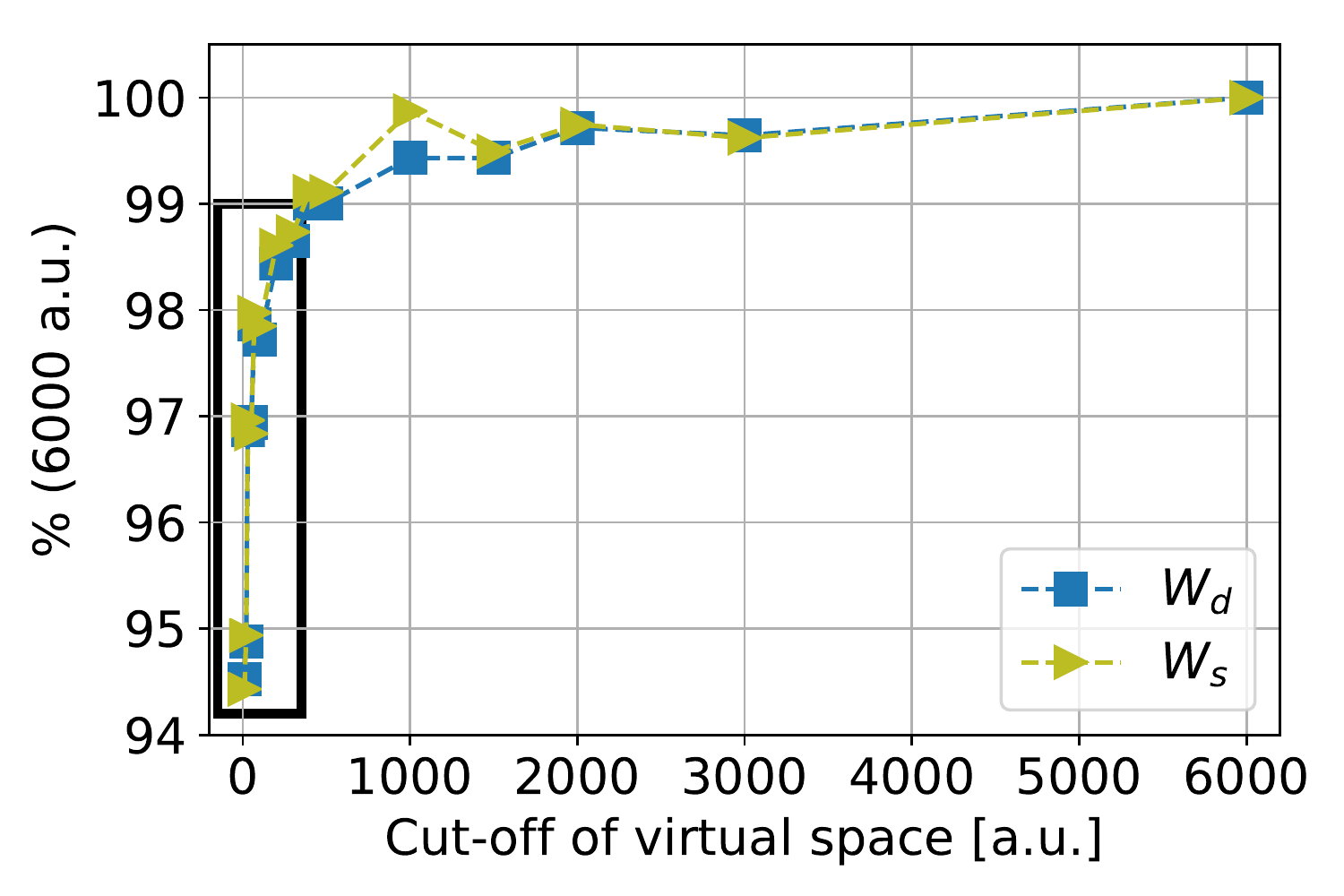}
\includegraphics[width=\linewidth]{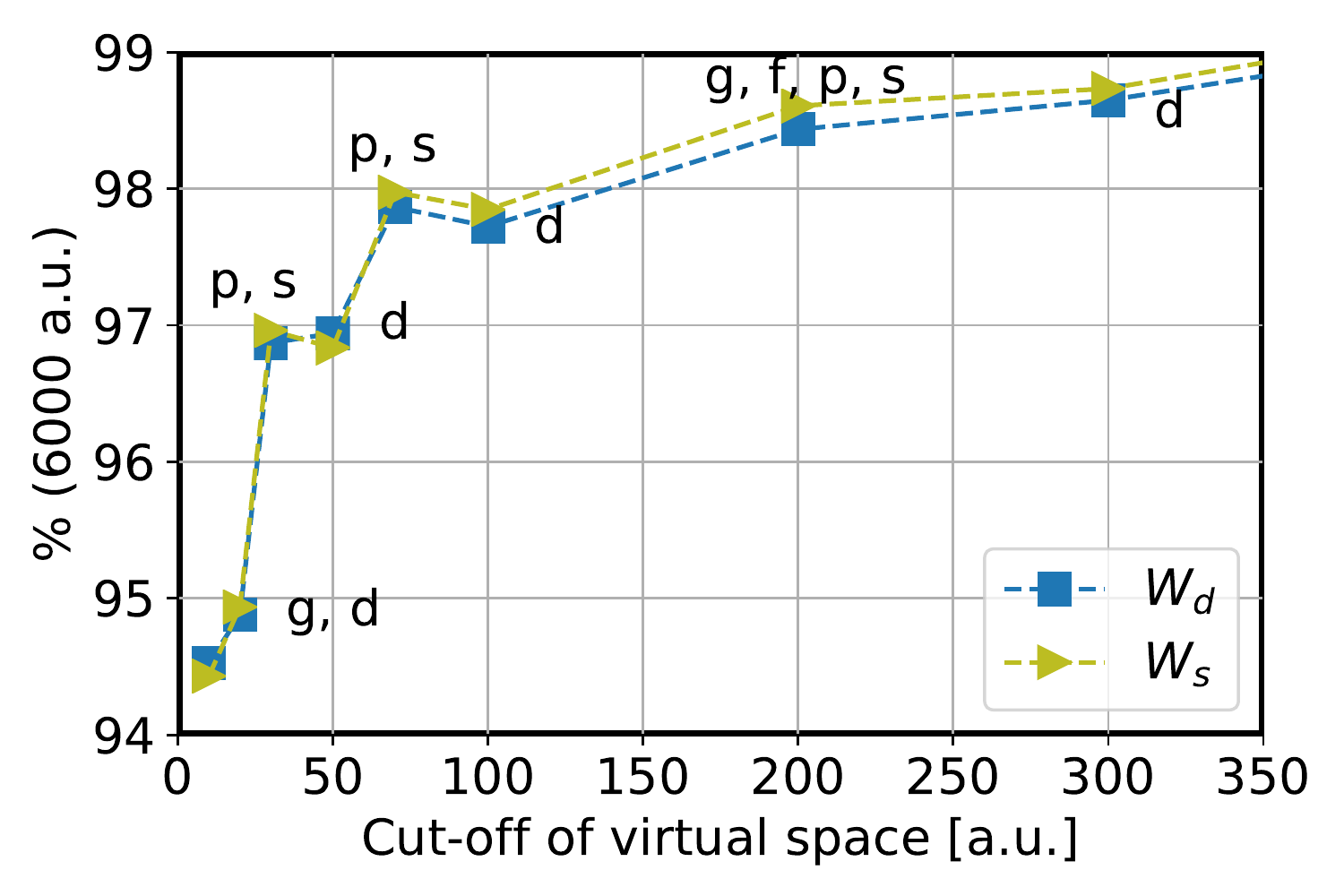}
\caption{Top: Dependence of calculated $W_d$ and $W_s$ on the virtual space cut-off, in percent, relative to the results obtained with a cut-off of 6000 a.u. Bottom: zoom-in of the area indicated by the black rectangle on the upper panel including specification of Ba orbitals added at each point. The presented results were obtained on the CCSD(T)/aetz level. See also Supplementary Information, Tab. S4.} \label{fig:virt_space}
\end{figure}

When using large basis sets, such as in the present study, it can be advantageous to limit the number of virtual orbitals included in the correlation treatment in order to reduce computational costs. In the following, we therefore perform investigations to determine a reasonable virtual space cut-off that will introduce only minor errors.

In Fig. \ref{fig:virt_space} the effect of cutting off virtual orbitals at various energies on the calculated $W_d$ and $W_s$ is shown, relative to a cut-off of 6000 a.u., above which additional functions are expected to have a minor effect. The two properties behave very similarly in regard to the changes in the size of the virtual space, which is also consistent with our findings in the previous section concerning the effect of the number of correlated electrons. The only discrepancy is found at a cut-off of 1000 a.u. where the $W_s$ value increases more than the $W_d$ value. For a cut-off of 500 a.u. both properties are underestimated by $\sim$ 1\% and for a cut-off of 2000 a.u. this becomes $\sim$ 0.3\%. We therefore chose to use a cut-off of 2000 a.u. for the results presented in the rest of this paper.

In the zoom in of Fig. \ref{fig:virt_space}, it can be seen that some virtual Ba orbitals contribute more than others, depending on their type. In particular, Ba s and p orbitals seem to contribute the most. It is also interesting to note that between the fifth and the sixth points both properties decrease due to the addition of the d-functions of Ba to the correlation space. In a similar fashion, adding tight d-functions to the basis set led to a decrease in the value of the $W_d$ parameter (Section \ref{sec:basis}). Since the virtual orbitals situated on the F atom are expected to have a negligible contribution, we do not address these in our discussion.

\subsubsection{Perturbative triples}\label{sec:corr}

\begin{figure}
    \centering
    \includegraphics[width=\linewidth]{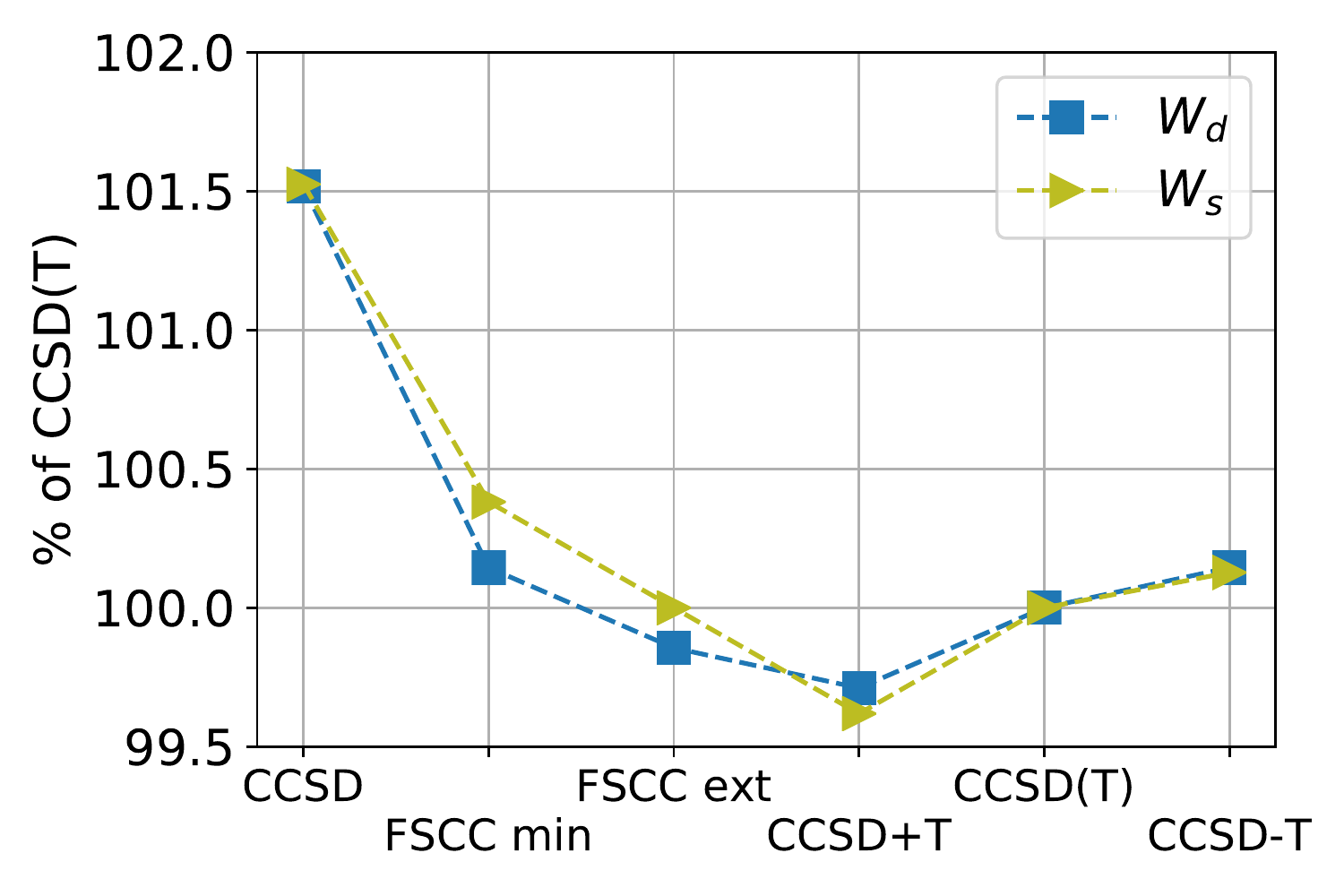}
    \caption{Effect of treatment of electron correlation on the calculated $W_d$ and $W_s$, in percent, relative to the recommended CCSD(T) results. The cvqz basis set was used in the calculations. See also Table \ref{tab:tests}. }
    \label{fig:corr}
\end{figure}

In this section we investigate the effect of the excitation rank and estimate the magnitude of the error that is introduced by truncating the coupled cluster expansion. Table \ref{tab:tests} and Fig. \ref{fig:corr} present the results obtained on different levels of treating electron correlation.

The effect of correlation is very similar for the two properties, with CCSD results higher by $\sim$ 27\% compared to the DHF values; the pertubative triples lower the $W_d$ and $W_s$ values by $\sim$ 1.5\%. The spread in the contributions of the different schemes for treatment of perturbative triple excitations (CCSD+T, CCSD(T), and CCSD$-$T) is around 1/3 of the total effect of triples compared to CCSD, as can be seen in Fig. \ref{fig:corr}. We can therefore conclude that the perturbative triple excitations are stable for these two properties (in contrast to the case of the hyperfine structure constants\cite{Haase2020a}) and that the results are reasonably converged with respect to excitation rank.  

Another test of higher order correlation effects and effects related to any multi-reference character is by comparison of FSCC to the single reference results. In Table \ref{tab:tests} FSCC results are shown using a minimum ($min$) and extended ($ext$) model space in which the valence orbital ($\sigma$) and additional 5 lowest virtual orbitals ($\pi$, $\pi$, $\delta$, $\delta$, $\sigma$) were included, respectively. The results are similar to the CCSD(T) results with FSCC $min$ being in both cases slightly higher than CCSD(T) and FSCC $ext$ slightly lower than CCSD(T) in the case of $W_d$ and practically identical to CCSD(T) in the case of $W_s$. This indicates that the single-reference description used in the CCSD(T) method is indeed suitable, as expected for the BaF $X^2\Sigma_{1/2}^+$ ground state.

For the recommended values we choose to use the CCSD(T) method since the difference with respect to the formally more correct CCSD$-$T method is very small and since the CCSD(T) method is well-known and widely used, facilitating the comparison with other works.

\subsection{Treatment of relativity and nuclear model}\label{sec:rel}

Table \ref{tab:tests} also contains the comparison of the results obtained using different description of relativistic and related effects on the CCSD(T)/cvqz level.

So far we have used the Dirac-Coulomb (DC) Hamiltonian, Eq. (\ref{eq:DC}), in which the one-electron part is described by the 4-component Dirac-Hamiltonian and the 2-electron interaction by the Coulomb operator.

\begin{table}[t]
\caption{$W_d$ [$10^{24}\frac{\text{Hz}}{e\cdot \text{cm}}$] and $W_s$ [kHz] obtained using different treatment of electron correlation, relativity and nuclear charge description (the cvqz basis set was used in the calculations).}\label{tab:tests}
\begin{tabular}{llcc}
\hline
\multicolumn{2}{c}{method} & $W_d$ [$10^{24}\frac{\text{Hz}}{e\cdot \text{cm}}$] & W$_s$ [kHz] \\
\hline
\multicolumn{3}{l}{Correlation} \\
~~ & DHF             & 2.332 & 6.215  \\
 & DC-MP2            & 2.884 & 7.668  \\
 & DC-CCSD           & 3.180 & 8.416  \\
 & DC-FSCC (0,1) $min$ & 3.137 & 8.321  \\
 & DC-FSCC (0,1) $ext$ & 3.128 & 8.290  \\
 & DC-CCSD+T         & 3.123 & 8.258  \\
 & DC-CCSD(T)        & 3.132 & 8.290  \\
 & DC-CCSD$-$T       & 3.137 & 8.300  \\
\multicolumn{3}{l}{Treatment of relativity} \\
 & X2C-CCSD(T)       & 3.121 & 8.279 \\
 & \textcolor{blue}{DC-CCSD(T)+$\Delta$Gaunt(DHF)}  & 3.076 & 8.258 \\
\multicolumn{3}{l}{Nuclear model} \\
 & Point nucleus (DC-CCSD(T))          & 3.177 & -- \\
\hline
\end{tabular}
\end{table}
A popular alternative to the computationally expensive 4-component approach is the exact 2-component (X2C) method\cite{Ilias2007} in which the large and the small component are exactly decoupled and the positive energy spectrum of the 4-component Hamiltonian is reproduced within numerical precision\cite{Saue2012a}. In Table \ref{tab:tests} we show the X2C results for $W_d$ and $W_s$. The deviation from the 4-component value in the case of $W_s$ is indeed very small (0.1\%) whereas the deviation for $W_d$ is slightly larger, namely 0.4\%. Similar deviations were observed for other properties such as HFS constants\cite{Haase2020a}, parity-violating matrix elements\cite{Bast2011} and contact densities\cite{Knecht2011}.

The Coulomb operator can be considered as a non-relativistic description of the 2-electron interaction since it does not take the finite speed of light into account. In other words, this interaction is instantaneous. For a first-principles relativistic description of the 2-electron interaction one has to turn to the theory of quantum electrodynamics (QED). The lowest order of QED radiative correction consists of single virtual transverse polarized photon exchange between the interacting electrons. In the Feynman gauge these corrections can be formulated in terms of a magnetic interaction (Gaunt term) and in the Coulomb gauge it includes also the additional retardation effect (Breit term)\cite{Kutzelnigg1987,Kutzelnigg2012a}. In the DIRAC17 program, the Gaunt interaction is available at the DHF level\cite{Pernpointner2002} and we tested its effect on the calculated $W_d$ and $W_s$.

The results in Table \ref{tab:tests} show that this interaction has a different effect on the two properties. Similar to the case of the X2C results, the effect for $W_d$ is larger ($\sim$ -1.8\%) than the effect on $W_s$ ($\sim$ -0.4\%). Similar magnitude of this contributions was observed in Ref. \citenum{Talukdar2020}. In both cases the effect is significantly larger than in the case of the HFS constants ($\sim$ -0.04\%)\cite{Haase2020a} which could be explained by the fact that $W_d$ and $W_s$ are \textcolor{blue}{highly} relativistic properties and that they are therefore more sensitive to the treatment of relativity. 

It should, however, be noted that the current implementation of the Gaunt interaction lacks contributions from a) relaxation effects (due to the use of Kramer's restricted DHF orbitals), b) correlation effects \textcolor{blue}{and c) terms of the order $\alpha^2$ originating from the full Breit interaction missing both in the Gaunt interaction itself and in the expression for the eEDM Hamiltonian, Eq. (\ref{eq:eEDM}).} In the case of HFS constants we found the calculated Gaunt contribution to be too low and with the opposite sign compared to the more sophisticated methods\cite{Haase2020a}. However, to our knowledge, no study of the Gaunt and/or Breit contributions to these properties has yet been reported on a higher level of theory and we consequently choose to include the present result as an order of magnitude estimate for the actual effect.

%\subsection{Nuclear model}

In relativistic electronic structure methods, it is common practice to model the nuclear charge using either a Gaussian or a Fermi charge distribution\cite{Visscher1997b}.
To investigate the error introduced by this approximation, we compare the results obtained using the Gaussian model, as implemented in the DIRAC17 program to the results of calculations that use a point charge description (Table \ref{tab:tests}). We do not consider the effect of using the point charge model on the calculated $W_s$, as one would also need to replace the nuclear charge distribution contained in the operator, Eq. (\ref{eq:ws}), which would induce a too drastic change in the definition of this operator. 
%, we don't consider this effect on $W_s$. \pabh{Reasoning could be clearer here.}

For $W_d$ the observed effect is 1.5\%. Since the point charge model is a very crude approximation, the error introduced by using a Gaussian charge distribution can be expected to be at least an order of magnitude smaller than the difference observed here. Therefore, we neglect this effect in the determination of the uncertainty of the predicted $W_s$ and $W_d$ values.  In Ref. \citenum{Skripnikov2015} the uncertainty associated with modelling the nucleus using the Fermi charge distribution was estimated to be around 1\% for $W_d$ in ThO. This larger uncertainty is likely due to the much larger $Z$ of Th. 
%as the dependence of nuclear description increases with $Z$\cite{Visscher1997b}. 

\begin{table}[b]
\renewcommand*{\arraystretch}{1.25}
    \caption{ \textcolor{purple}{$W_d$ and $W_s$ in the three lowest vibrational levels compared to the values at the equilibrium distance, $R_e$, determined from the fit of data points shown in Fig. \ref{fig:geometry}. The DC-CCSD(T)/aetz method was used.} } \label{tab:vib}
    \centering
    \begin{tabular}{lcccc}
    \hline
    Vib. level & $W_d$ [$10^{24}\frac{\text{Hz}}{e\cdot \text{cm}}$] & \%($R_e$) &  $W_{s}$ [kHz] & \%($R_e$) \\
    \hline
    $R_e$ (2.178\AA) & 3.201 & -    & 8.375 & - \\
    $v_0$            & 3.202 & 0.03 & 8.387 & 0.15 \\
    $v_1$            & 3.214 & 0.43 & 8.423 & 0.58 \\
    $v_2$            & 3.227 & 0.83 & 8.455 & 0.96 \\
    \hline
    \end{tabular}
\end{table}

\subsection{\textcolor{purple}{Vibrational effects}}

\begin{figure}[t]
    \centering
    \includegraphics[width=\linewidth]{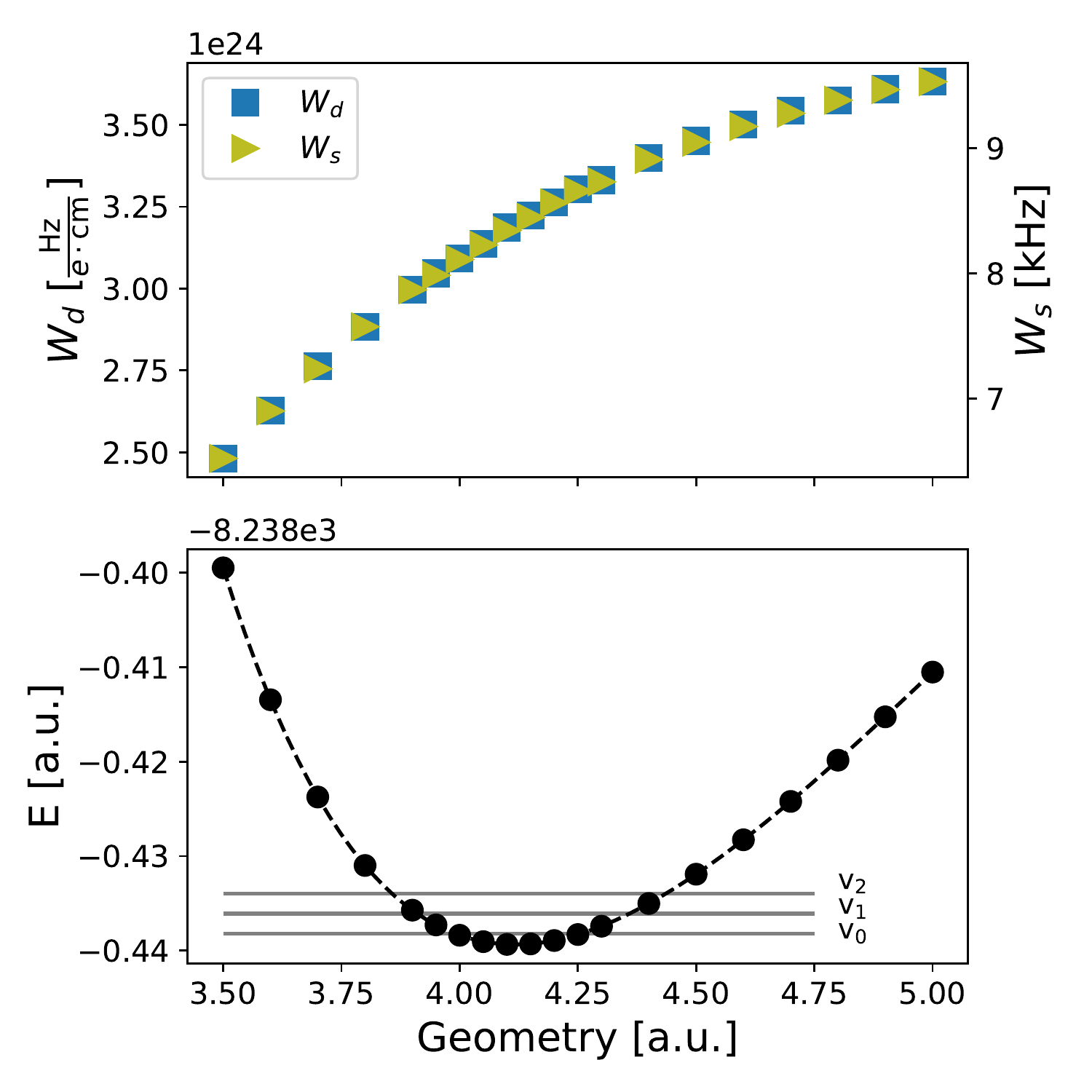}
    \caption{ \textcolor{purple}{Top: $W_d$ (left axis) and $W_s$ (right axis) calculated at different geometries at the DC-CCSD(T)/aetz level. Bottom: Corresponding unperturbed energies. Dashed line represents the fitted potential energy curve and solid lines the three lowest vibrational levels.} }
    \label{fig:geometry} 
\end{figure}

\textcolor{purple}{Before turning to the discussion of the theoretical uncertainty we need to consider the effects of vibrations on the $W_d$ and $W_s$ parameters. For this analysis we used the VIBROT module of OpenMolcas\cite{Fdez.Galvan2019} which fits a given potential energy surface (PES) as well as a property surface. Based on the fitted PES, the vibrational levels and corresponding wave functions are determined by solving the vibrational Schr\"odinger equation numerically with the aid of the Numerov-Cooley approach\cite{Ingamells2000}. Finally, the given property is determined for each vibrational level as the expectation value over the corresponding vibrational wave function. The single points for a range of geometries were calculated at the DC-CCSD(T)/aetz level of theory and the results as well as the three lowest vibrational levels are shown in Fig. \ref{fig:geometry} and Supplementary Information Tab. S5. Note the very similar dependence of the $W_d$ and $W_s$ parameters on the internuclear distance. }

\textcolor{purple}{The vibrational corrections to $W_d$ and $W_s$ in the three lowest vibrational levels are shown in Tab. \ref{tab:vib}, calculated as the difference between the vibrationally averaged property for each vibrational level and the property calculated at $R_e$. The resulting corrections for the considered levels are small due to the fact that the property curve is close to being linear in this region. Combined with the symmetric shape of the PES in the low vibrational levels means that the observed changes in $W_d$ and $W_s$ are integrated out. The results of this analysis will be used in the estimation of the theoretical uncertainty as presented in the following section. }

\subsection{Uncertainty estimation}\label{sec:error}

We choose to use the results obtained on the DC CCSD(T) level of theory using the cvqz basis set, correlating all electrons, and cutting of the virtual space at 2000 a.u. as the recommended values. We thus arrive at values of 3.13$ \times 10^{24}\frac{\text{Hz}}{e\cdot \text{cm}}$ for $W_d$ and 8.29 kHz for $W_s$.

\begin{figure}[t]
    \centering
    \includegraphics[width=\linewidth]{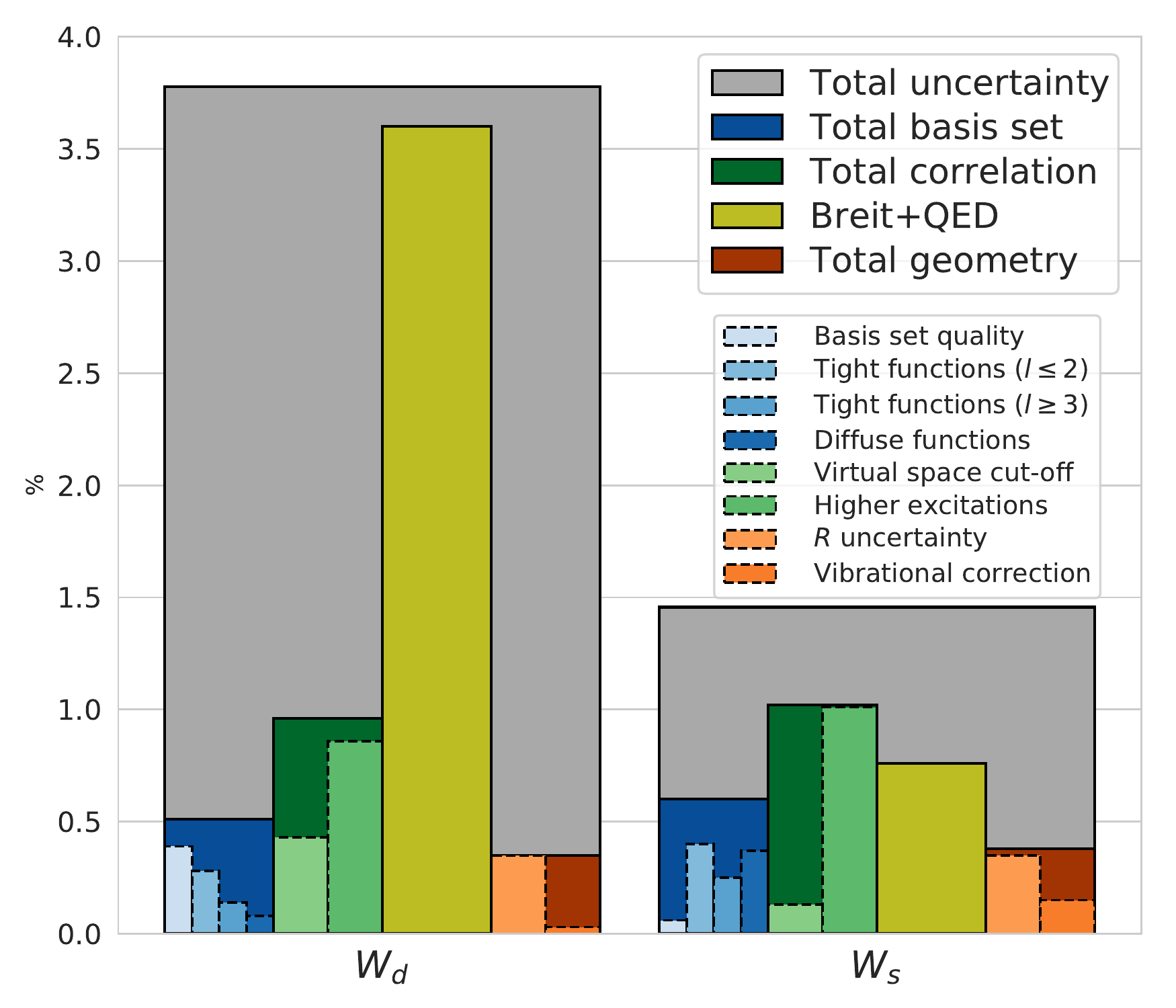}
    \caption{Graphical representation of the total uncertainty as well as the individual contributions in percent relative to the recommended values (DC-CCSD(T)/cvqz level). Also the total contributions from the basis set and correlation effects are shown. }
    \label{fig:unc}
\end{figure}

Based on the exhaustive analysis of various computational approximations presented above, we can determine the uncertainty of the calculated $W_d$ and $W_s$ enhancement factors. We follow the strategy employed in earlier publications dealing with similar properties as well as with HFS constants\cite{Hao2018,Denis2019,Denis2020,Haase2020a}, where the uncertainty associated with each computational approximation is estimated separately in a systematic fashion. % and presented in an easy-to-follow way.
These individual uncertainties are shown in Table \ref{tab:unc} and illustrated in Fig. \ref{fig:unc}. In the following we address the main conclusions from this analysis; the reader is referred to the previous sections for a thorough discussion of the observed effects. The scheme we use to determine the individual uncertainties is constructed to provide realistic yet conservative estimates. For some properties the sign of the estimated uncertainty is not obvious and we therefore choose to consider the absolute values in all cases. 

\begin{table}[t]
\caption{Summary of the sources of uncertainty (absolute values) of the calculated $W_d$ [$10^{24}\frac{\text{Hz}}{e\cdot \text{cm}}$] and $W_s$ [kHz] enhancement parameters in BaF along with the absolute total uncertainty and in percent relative to the recommended  DC CCSD(T)/cvqz  value.}
\begin{tabular}{llcc}
\hline
 \multicolumn{1}{l}{Source} & \multicolumn{1}{l}{Estimation scheme}  & $\delta W_d$ & $\delta W_s$ \\
 \hline
\multicolumn{4}{l}{Basis set}  \\ 
~ Quality & (vqz - vtz)/2                          & 0.012 & 0.005  \\ 
~ Diffuse funct. & s-aug-vqz - vqz               & 0.002 & 0.031  \\ 
~ Tight funct. $l\leq2$ & (s, p, d)              & 0.009 & 0.033  \\
~ Tight funct. $l\geq3$ & aeqz - cvqz            & 0.004 & 0.021  \\ 
\multicolumn{4}{l}{Correlation}                          \\ 
~ Virtual space cut-off& 6000 a.u - 2000 a.u.    & 0.014 & 0.011  \\ 
~ Higher excitations & (CCSD$-$T - CCSD+T)$\cdot$2 & 0.027 & 0.084  \\  
\multicolumn{4}{l}{Relativity}                           \\
~ Breit+QED   & (DC\textcolor{purple}{$+\Delta$}G - DC)$\cdot$2               & 0.113 & 0.064 \\ 
\multicolumn{4}{l}{\textcolor{purple}{Geometry}} \\
%~ \textcolor{purple}{$R_{\text{exp}}$ uncertainty    } & $R_{\text{exp}}$ - ($R_{\text{exp}}$ - $\delta R_{\text{exp}}$) (aetz) & 0.011 & 0.029 \\
~ \textcolor{purple}{$R$ uncertainty    } & $R$ - ($R$ - $\delta R$) (aetz) & 0.011 & 0.029 \\
~ \textcolor{purple}{Vibrational effects} & $R_e$ - $v_0$ (aetz)                  & 0.001 & 0.013 \\
\cline{3-4}
\multicolumn{4}{l}{Total}  \\                       
~ sum & $\sqrt{\Sigma_i \delta_i^2}$             & 0.11\textcolor{purple}{9} & 0.1\textcolor{purple}{22} \\ 
~ \%                             &               & 3.7\textcolor{purple}{9} & 1.4\textcolor{purple}{7} \\ 
 \hline
\end{tabular}
\label{tab:unc}
\end{table}

\paragraph{Basis set}

The uncertainty stemming from using a finite rather than a complete basis set can be determined as a combination of three separate parameters: 1) general basis set quality, determined as half the difference between the vqz and the vtz results, 2) the quality of the description of the valence region (the amount of diffuse functions), estimated as the difference between the s-aug-vqz and vqz results and 3) the quality of the description of the core region (the amount of tight functions). The latter uncertainty can be additionally split into uncertainty due to functions with low (s, p and d) and high (f, g and h) angular momentum, taken as the effect of additional tight s, p and d functions shown in Fig. \ref{fig:tight} and as the difference between the aeqz and cvqz results, respectively. For $W_d$, the uncertainty stemming from the basis set quality dominates, whereas for $W_s$ the missing tight s, p and d- and diffuse functions have the largest effect (as for this property we observed much more smooth convergence with respect to the basis set quality than for $W_d$). In both cases, the uncertainty introduced by the incompleteness of the basis set amounts to $\sim$ 0.5\%. 

\paragraph{Electron correlation}

Two aspects contribute to the uncertainty associated with the correlation treatment, namely the cut-off of virtual orbitals and the higher rank excitations. Ideally, one would study the convergence with respect to the excitation rank by comparing the CCSD and CCSDT methods. However, calculations with the latter method are not presently feasible and instead we take twice the spread in the values obtained with the different schemes for including perturbative triples. Another possibility to estimate the effect of the residual triple and the higher excitation is by comparing CCSD(T) with CCSD results, which in fact gives a similar uncertainty. The observed conservative effect of $\sim$1\% turns out to be very similar for the two properties. For $W_s$ this is the leading contribution to the total uncertainty. In Ref. \citenum{Skripnikov2015} the effect of full triple and perturbative quadruple excitations (relative to CCSD(T)) was found to be $\sim$ 0.1\% for $W_d$ in ThO, i.e. considerably smaller than the present estimate. In Ref. \citenum{Talukdar2020}, the uncertainty due to higher rank excitations was assumed to be 3.5\% which is more conservative than determined here. The cut-off of the virtual space is a considerably smaller source of uncertainty and is estimated as the difference between a cut-off of 6000 and 2000 a.u.

\paragraph{Relativistic effects}

Here we consider the error introduced by describing the two-electron interaction by the non-relativistic Coulomb potential and by neglecting the higher order QED corrections. This is done by considering the effect of the Gaunt interaction, as discussed in Sec. \ref{sec:rel}. Since QED corrections to properties can be of similar magnitude as the Breit interaction (as for example in the case of HFS constants\cite{Ginges2017}), we multiply the effect of Gaunt by 2 to give the uncertainty estimate. For $W_d$ the uncertainty due to higher order relativistic effects dominates the total uncertainty estimate with $\sim$ 3.6\%, whereas for $W_s$ this effect is considerably smaller, namely $\sim$ 0.76\%.

\paragraph{\textcolor{purple}{Geometry}}

\textcolor{purple}{In the following we investigate the uncertainty stemming from the employed geometry. %using CCSD(T) and the aetz basis set. 
In all the calculations presented so far, the experimental bond length of 2.162\AA~was used. This bond length was measured with a precision of 0.007\AA\cite{NISTBaF,Knight1971} and we estimate the corresponding uncertainty in $W_d$ and $W_s$ using the fitted property curves presented in Fig. \ref{fig:geometry} and Supplementary Information Tab. S5. The fitted curves were used to determine $W_d$ and $W_s$ both at the employed bond length of 2.162\AA~and including the uncertainty, i.e. 2.162 $\pm$ 0.007\AA~and the results are shown in Supplementary Information Tab. S6. The largest absolute difference is 0.36\% for both $W_d$ and $W_s$.}

\textcolor{purple}{Another source of uncertainty related to the geometry is that of vibrational effects. The planned EDM experiment will use various slowing and cooling techniques to obtain a slow intense beam of BaF molecules in the ground electronic, vibrational and rotational state\cite{Aggarwal2018}. Consequently, we only need to consider the vibrational correction in the $v_0$ level for the uncertainty estimate. These results are shown in Tab. \ref{tab:vib} and we obtain an uncertainty contribution of 0.03\% and 0.15\% for $W_d$ and $W_s$, respectively. The observed effect is similar to that reported by Skripnikov et al. for $W_d$ in ThO\cite{Skripnikov2015}. As can be seen from Fig. \ref{fig:unc}, the uncertainty related to the geometry provides the smallest contribution to the total uncertainty.} 

\paragraph{Total uncertainty}

To obtain the total uncertainty, we add the individual sources of uncertainty in a quadratic manner, which assumes the different contributions to be independent. Although effects such as basis set size and electron correlation are known to be strongly interdependent, this assumption can be considered suitable as long as we are dealing with higher order corrections. %In our analysis we have neglected the uncertainty due to any vibrational effects on $W_d$ and $W_s$, as we expect these to be very small. For example,  in Ref. \citenum{Skripnikov2015}, Skripnikov et al. found a small effect of 0.12\% on $W_d$ in ThO.

The total uncertainty is almost twice as large for $W_d$ compared to $W_s$ due to the significant Gaunt contribution. This highlights the need for a separate computational investigation for different molecules and even different properties in the same system, in order to determine both most suitable computational scheme and to estimate the uncertainties. 

The total uncertainties for $W_d$ and $W_s$ of 3.8\% and 1.5\% presented here are smaller than previously reported for calculations based on similar methods. In Ref. \citenum{Skripnikov2015}, an uncertainty of 7\% on $W_d$ of ThO was estimated using the 2-component CCSD(T) method. In Ref. \citenum{Sasmal2015a}, the CCSD approach was used to calculate $W_d$ in PbF and an uncertainty of 4\% was suggested. Considering studies of $W_d$ in BaF, no error bars were given in Ref. \citenum{Abe2018}, which showed results on the CCSD level, whereas in Ref. \citenum{Talukdar2020}, the uncertainty was estimated to be 8\% using the same computational approach.

In order to test whether our estimated uncertainty is realistic, we followed the usual strategy where we used the same computational method to calculate the $^{137}$Ba HFS constants in BaF and compared the results to accurate experimental data. Using HFS constants to benchmark methods for calculating $P,T$-odd parameters relies on the fact that both types of properties are sensitive to the quality of the description of the wave function in the vicinity of the nucleus. This has in fact enabled semi-empirical extraction of $W_d$ and $W_s$ from the experimentally measured HFS constants\cite{Kozlov1985,Kozlov1995}.  For the HFS constants we found a deviation of 0.3\%\cite{Haase2020a} between theory and experiment, which provides a conformation that the low uncertainty estimate presented in this work is indeed reliable. 

\subsection{Comparison with previous results}

\begin{table}[t]
%\mytablefontsize
%\renewcommand*{\arraystretch}{\myarraystrech}
\caption{Previous predictions for the $W_d$ [$10^{24}\frac{\text{Hz}}{e\cdot \text{cm}}$] and $W_s$ [kHz] parameters in BaF as well as the current recommended values with associated uncertainty. Numbers in parenthesis show the estimated uncertainty when given.} \label{tab:comparison}
\begin{tabular}{llccc}
\hline
%                     &  Author               & Method            &  \eeff  & $W_s$  \\  
Ref.                                & Method                           & $W_d$  & $W_s$  \\  
\hline 
\citenum{Kozlov1995, Knight1971}    & HFS + SE\footnotemark[1]         & 3.5  (20\%)  & 11 (20\%)  \\
\citenum{Kozlov1995,Ryzlewicz1980}  & HFS + SE\footnotemark[1]         & 4.1  (20\%)  & 13 (20\%)  \\
\citenum{Kozlov1997}                & GRECP RASSCF\footnotemark[2]     & 2.2          & 5.9  \\
                                    & GRECP RASSCF-EO\footnotemark[2]  & 3.6          &      \\
\citenum{Nayak2006,Nayak2007}       & 4c RASCI\footnotemark[3]         & 3.52         &  9.7 \\
\citenum{Meyer2008}                 & NR + MRCI\footnotemark[4]        & 3.0          &      \\
\citenum{Fukuda2016}                & 4c CISD\footnotemark[5]          & 2.09         &      \\
\citenum{Gaul2017,Gaul2019}         & GHF-ZORA\footnotemark[6]         & 3.32 (20\%)  & 8.67 (20\%)  \\
                                 & GKS-ZORA/B3LYP\footnotemark[6]   & 2.90 (20\%)  & 7.58 (20\%)  \\
\citenum{Abe2018}                   & 4c LE CCSD\footnotemark[7]       & 3.14         &      \\
                                    & 4c FF CCSD\footnotemark[8]       & 3.12         &      \\
\citenum{Sunaga2018}                & 4c LE CCSD\footnotemark[7]       & 3.20         & 8.4  \\
\citenum{Talukdar2020}              & 4c Z-vector CCSD\footnotemark[9] & 3.15 (8\%)   & 8.35 (8\%) \\ \noalign{\smallskip}
This work                        & 4c FF CCSD(T)                    & 3.13 \textcolor{blue}{(3.8\%)} & 8.29 \textcolor{blue}{(1.5\%)} \\
\hline
\end{tabular}
\footnotetext[1]{Semi-empirical (SE) method based on the HFS parameters of Knight et al\cite{Knight1971} and Ryzlewicz et al\cite{Ryzlewicz1980}.}
\footnotetext[2]{Generalized relativistic effective core potentials (GRECP) restricted active space self-consistent field method (RASSCF) with and without an effective operator (EO)}
\footnotetext[3]{4-component (4c) restricted active space configuration interaction (RASCI)}
\footnotetext[4]{Non-relativistic (NR) multi-reference configuration interaction (MRCI) with effective core potentials (ECP)}
\footnotetext[5]{4c configuration interaction with single and double excitations (CISD)}
\footnotetext[6]{Generalized Hartree Fock (GHF) and generalized Kohn Sham (GKS) with the zeroth-order regular approximation (ZORA)}
\footnotetext[7]{CCSD with only linear terms in the expression for the expectation value.}
\footnotetext[8]{CCSD in combination with the finite field (FF) method.}
\footnotetext[9]{CCSD in combination with the Z-vector method.}
\end{table}

Several authors have investigated $W_d$, and, to a lesser extent, $W_s$, in BaF over the past decades and on various levels of theory\cite{Kozlov1995, Kozlov1997, Kozlov1997, Nayak2006, Nayak2007, Meyer2008, Fukuda2016, Gaul2017, Gaul2019, Abe2018, Sunaga2018, Talukdar2020}. In Table \ref{tab:comparison}, a comparison of our results with previous studies is shown; the results at the highest level of theory are presented from each study. We have converted results presented as \eeff to $W_d$ notation using the relation $W_d = \frac{1}{\Omega} \eeffm$, where $\Omega=1/2$ in the case of the BaF $X^2\Sigma_{1/2}^+$ ground state. In addition to the variety of methods used, the different results were obtained with different basis sets and some using different geometries, which makes a direct comparison problematic. 

The relativistic CCSD method has recently become a widely used approach for accurate predictions of the $P,T$-odd properties and we see that Refs. \citenum{Abe2018}, \citenum{Sunaga2018} and \citenum{Talukdar2020} predict very similar values, all within the uncertainty of the present work, although the result of Ref. \citenum{Sunaga2018} can be expected to have a fortunate cancellation of errors as it was obtained with the vtz basis set and a relatively low virtual cut-off of 80 a.u. Generally, the majority of the CCSD results today are obtained using either the Z-vector method, where properties are calculated as expectation values, or the more approximative LE method, which uses a linearized expression of the CC expectation value. In Ref. \citenum{Abe2018}, the latter approximation was compared to the finite field method which showed deviations $<$1\%. The advantage of the present results compared to the previous CCSD studies is the inclusion of perturbative triples as well as the possibility of providing a transparent and reliable uncertainty estimate.

While a detailed comparison between the present and the earlier results obtained on a lower level of theory is not of particular interest here, a few remarks should be made. Gaul and co-workers calculated $W_d$ at the 2-component ZORA level using the general Hartree-Fock and Kohn-Sham methods\cite{Gaul2017}. Contrary to our findings in Sec. \ref{sec:corr}, a decrease in the calculated $W_d$ value was observed when including electron correlation compared to Hartree-Fock. Gaul et al. attributed this difference to the opposite effects of spin-polarization and electron correlation, since spin-polarization is already taken into account on the GHF level. This might also explain the relatively high RASSCF-EO result, since the effective operators are associated with the inclusion of spin-polarization. In our approach, we account for spin polarization only at the coupled cluster level starting from restricted Hartree-Fock orbitals; however, we expect that this is compensated in the coupled cluster procedure. 

The result of Ref. \citenum{Fukuda2016} obtained with the 4c CISD method is significantly lower than most other theoretical values. This large difference was attributed to aspects concerning correlation treatment. In addition, a slightly different geometry was used in that study.

\section{Conclusions}

The $P,T$-odd molecular enhancement factors $W_d$ and $W_s$ are crucial for the interpretation of precision experiments that search for physics beyond the Standard Model using molecules. 
These parameters cannot be determined experimentally and the \textcolor{purple}{purpose of} this work was to calculate them for the ground state of the BaF molecule with the highest possible accuracy and to provide a transparent and systematic estimate of the associated uncertainties.  

We have used the 4-component finite field CCSD(T) method and the cvqz basis set to provide the recommended values of 3.13 $\pm$ $0.12 \times 10^{24}\frac{\text{Hz}}{e\cdot \text{cm}}$ and 8.29 $\pm$ 0.12 kHz for $W_d$ and $W_s$, respectively.
The main contributions to the uncertainties stem from the higher order relativistic effects in the case of $W_d$ and from the higher order correlation effects in the case of $W_s$. This, along with the difference in relative uncertainty (3.8\% and 1.4\% for $W_d$ and $W_s$, respectively), reflects the different dependence of the two properties on the various computational parameters. Whereas $W_d$ seems to be mostly sensitive to the description of the nuclear region and the relativistic effects, $W_s$ also shows dependence on diffuse functions included in the basis set, indicating its sensitivity to the description of the valence region. These differences highlight the need for separate in-depth computational studies for the different properties in order to provide high accuracy predictions and to assign reliable uncertainties. 

The predictions presented here will be used as part of the future interpretation of the NL-\textit{e}EDM experiment. In order to disentangle the effects stemming from the eEDM and the S-PS interaction, at least two measurements on systems for which the ratio $W_d/W_s$ differs significantly are required. Since this ratio depends mainly on $Z$, measurements on systems with a large difference in $Z$ are to be preferred\cite{SunAbePra19,Gaul2019}. The $W_d/W_s$ ratios for systems used in a few ongoing EDM measurements\cite{Aggarwal2018, Hudson2002, Cairncross2017, AndAngDem18} are listed in Table \ref{tab:ratio}. The results show that the largest difference is found between the lightest (BaF) and the heaviest (ThO) systems, whereas the ratios for YbF and HfF$^+$ are very similar due to the similar $Z$. Consequently, a precise measurement of the $P,T$-violating effects in BaF, combined with the ongoing measurements on the heavier systems, will be important in order to disentangle the two $P,T$-odd sources.

\begin{table}[t]
    \caption{Ratio $W_d/W_s$ [$10^{20}\frac{1}{e\cdot\text{cm}}$] of diatomic molecules currently being used in high precision EDM experiments.}
    \centering
    \label{tab:ratio}
    \begin{tabular}{lcc}
    \hline
        Ref. & System~ & $W_d/W_s$   \\
    \hline
        this work           & BaF\phantom{$^+$}     & 3.78 \\
        \citenum{Gaul2019}  & YbF\phantom{$^+$}     & 2.80\\
        \citenum{Fleig2017} & HfF$^+$               & 2.75 \\
        \citenum{Fleig2017} & ThO\phantom{$^+$}     & 1.72 \\
    \hline
    \end{tabular}
\end{table}

\section*{Dedication}

We dedicate this article to Emmy Noether, whose mathematical theorems profoundly influenced physics by demonstrating that every symmetry results in an associated conservation law. Her work marks the start of the relevance of symmetry in physics and is one of the pillars of the modern gauge theories of particle physics, which is the motivation for our article.

\section*{Acknowledgements}

We would like to thank the Center for Information Technology of the University of Groningen for their support and for providing access to the Peregrine high performance computing cluster as well as M. G. Kozlov for helpful discussions.
The NL-\textit{e}EDM consortium receives program funding (EEDM-166) from the Netherlands Organisation for Scientific Research (NWO). M. Ilia\v{s} acknowledges the support of the  Scientific Grant Agency, Grant  No. VEGA 1/0562/20.

\section*{Data Availability}

The data that support the findings of this study are available from the corresponding author upon reasonable request.

%\AtNextBibliography{\footnotesize}
%\printbibliography

%\bibliography{Computational,Finite_field,eEDM,AB}
%\bibliography{output}

%\begin{filecontents}{output.bbl}
%\end{filecontents}

\end{document}